%% file: main.tex
\documentclass[a4paper]{article}
\usepackage[utf8]{inputenc}
\usepackage{color}
\usepackage{amsmath}
\usepackage{amsmath}
\usepackage{amssymb}
\usepackage{amsfonts}
\usepackage{multirow}
\usepackage{multicol}
\usepackage{authblk}

\usepackage{graphicx}
\usepackage{color}
\usepackage[colorinlistoftodos]{todonotes}
\usepackage{enumerate}
\usepackage{tikz,xcolor}
\usetikzlibrary{patterns, calc, positioning, shapes.geometric, arrows, fit, arrows.meta, backgrounds}
\usepackage{float}
\usepackage{wrapfig}
\usepackage{subcaption}
\usepackage{caption}
\usepackage{mathtools}
\usepackage{pgfplots}
\pgfplotsset{compat=1.10}
\usepackage{xcolor}
\usepackage{amsthm}
\usepackage[margin=2.5cm]{geometry}
\usepackage{hyperref}
\newcommand{\R}{\mathbb{R}}

\newcommand{\Z}{{\mathbb Z}}
\newcommand{\C}{{\mathbb C}}

\newcommand{\fvec}{\mathbf{f}}
\newcommand{\wvec}{\mathbf{w}}
\newcommand{\argmin}{\mathop{\mathrm{arg\,min}}}

\tikzset{
    module/.style={%
        draw, rounded corners,
        minimum width=#1,
        minimum height=7mm,
        font=\sffamily
        },
    module/.default=2cm,
    >=LaTeX
}

\title{Learning a microlocal prior \\for limited-angle tomography}

\author[1]{Siiri Rautio}
\author[1]{Rashmi Murthy}
\author[2]{Tatiana A.~Bubba}
\author[1]{\\Matti Lassas}
\author[1]{Samuli Siltanen}
\affil[1]{Department of Mathematics and Statistics, University of Helsinki}
\affil[2]{Department of Mathematical Sciences, University of Bath}
\begin{document}
\date{}

\maketitle

\begin{abstract}
     Limited-angle tomography is a highly ill-posed linear inverse problem. It arises in many applications, such as digital breast tomosynthesis. Reconstructions from limited-angle data typically suffer from severe stretching of features along the central direction of projections, leading to poor separation between slices perpendicular to the central direction. A new method is introduced, based on machine learning and geometry, producing an estimate for interfaces between regions of different X-ray attenuation. The estimate can be presented on top of the reconstruction, indicating more reliably the true form and extent of features. The method uses directional edge detection, implemented using complex wavelets and enhanced with morphological operations. By using machine learning, the visible part of the wavefront set is first extracted and then extended to the full domain, filling in the parts of the wavefront set that would otherwise be hidden due to the lack of measurement directions. 
\end{abstract}

\section{Introduction}

X-ray tomography aims to recover the attenuation coefficient inside a physical body from a collection of radiographs recorded along different directions of projection. After calibrating the data, one can model the inverse problem mathematically as recovering a non-negative function $f:\R^n\rightarrow \R$ from a collection of line integrals of $f$, where $n$ is 2 or 3.
This study focuses on limited-angle tomography, where the directions of the lines in the dataset are severely restricted. Such cases appear in many important applications of X-ray tomography, for example, Digital Breast Tomosynthesis (DBT) and weld inspection. 

\begin{figure}
\begin{picture}(320,150) 
\put(-10,0){\includegraphics[height=6cm]{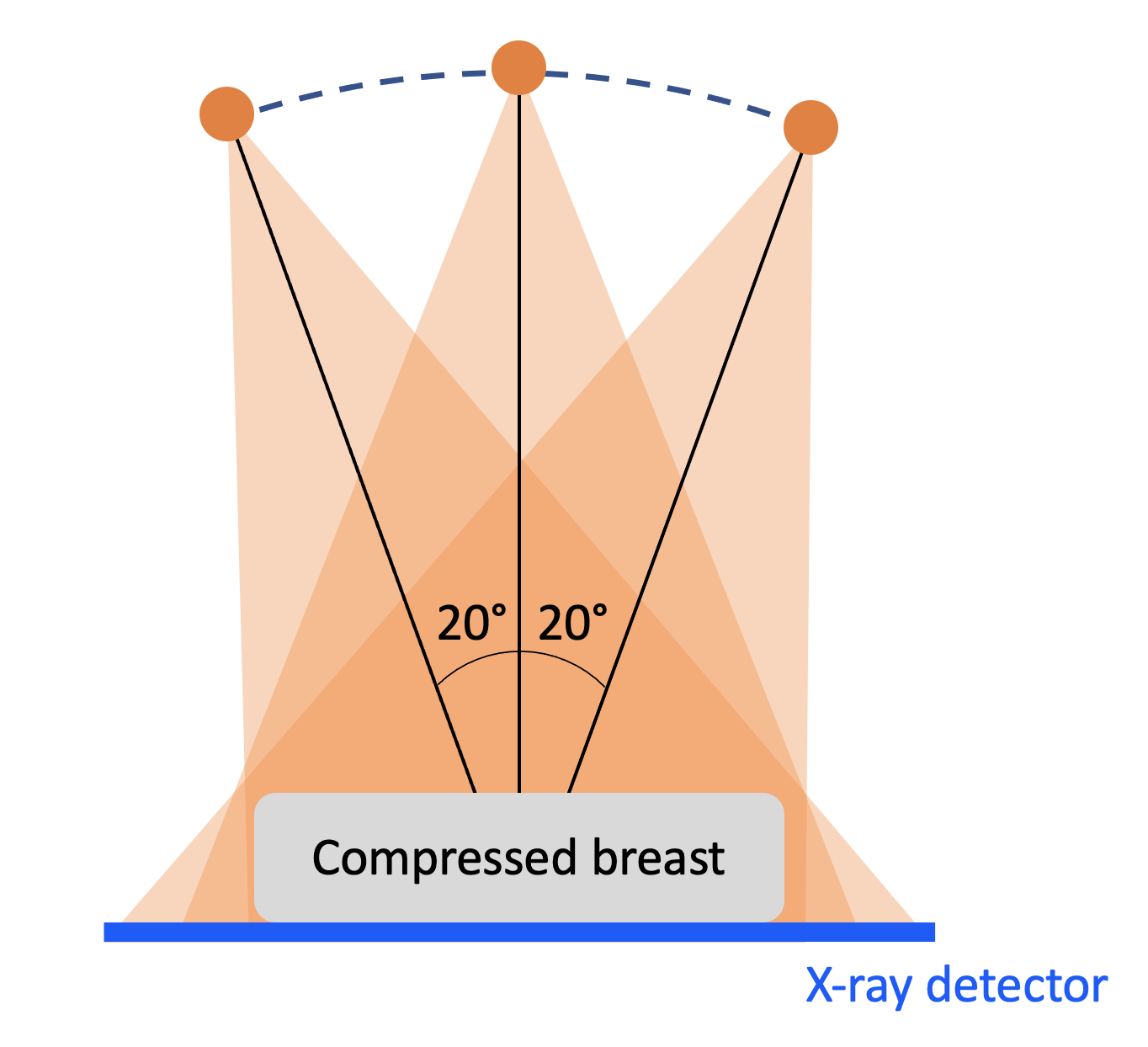}}
\put(-10,140){(a)}
\put(200,10){\includegraphics[height=5cm]{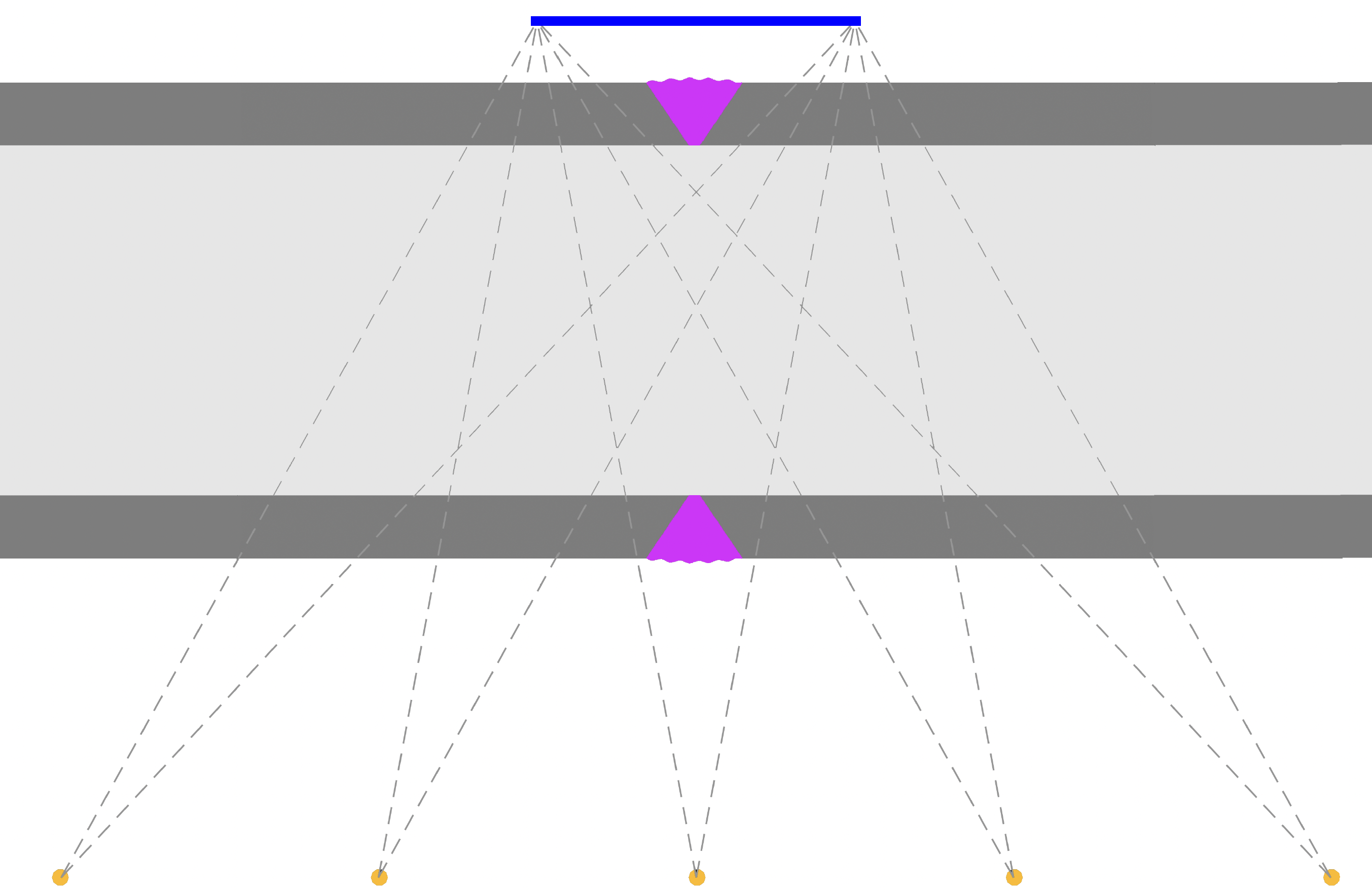}}
\put(180,140){(b)}
\end{picture}    
    \caption{
    (a) Imaging geometry of digital breast tomosynthesis, based on collecting several radiographs of the compressed breast at various angles. Typically, the most extreme directions are $\pm 20^\circ$ from the vertical, as shown in the image. While only three projections are depicted here, real DBT systems collect more (for example 15) images within the 40° angular range available. Limited-angle imaging geometry is forced by the compression of the breast against the detector plate. If the angle becomes too extreme, the rays passing through the tissue will not hit the detector. (b) Imaging geometry for X-ray inspection of weldings in pipelines used for example in process industry. Shown is a longitudinal cross-section of the pipe, with dark gray color indicating metal. The purple area is the weld connecting two pieces of pipe together. Limited-angle imaging geometry is necessary as both the X-ray source and the detector need to be outside the long pipeline. }
    \label{fig:DBTgeom}
\end{figure}

DBT is an enhanced form of mammography where one collects several X-ray images of a compressed breast within a limited angle of view. See Figure \ref{fig:DBTgeom}(a) for an illustration of the DBT measurement geometry. This provides more information about the three-dimensional structure of tissue and potentially leads to improved diagnosis of breast cancer \cite{dobbins2003digital,wu2004comparison,niklason1997digital,vedantham2015digital,rantala2006wavelet,piccolomini2016fast, landi2017limited, landi2019nonlinear}.

Long pipelines are used, for instance, in the process industry and the energy industry. They are often constructed by welding together several pieces of metal pipes. With welding, there is a risk of gas bubbles, cracks, or other defects forming inside the weld. One can use limited-angle X-ray tomography for non-destructive testing of the welds \cite{haith2017defect,silva2021x} with imaging geometry as shown in \ref{fig:DBTgeom}(b).

Limited-angle tomography is a very ill-posed inverse problem \cite{natterer1986computerized,davison1983ill,siltanen2003statistical}; in other words, it is very sensitive to modelling errors and measurement noise. In particular, reconstructions typically contain stretching artefacts along the central measurement direction and streak artefacts towards the extreme angle projections \cite{frikel2013characterization}. This makes it hard to recover the true form and extent of features. However, note that we know from \cite{quinto1993singularities} precisely which parts of the boundaries of objects are stably visible in limited-angle data. 

We introduce a new, partly data-driven and partly geometric method for recovering interfaces between different areas of X-ray attenuation in the target. We show how complex wavelets and neural networks can be used to first detect the visible parts of the interfaces and then extend them to the invisible parts. Before going into details of our algorithm, we review the related methodology in the literature.

In recent years, deep learning has been used to improve the quality of computed tomography (CT) reconstructions in different ways \cite{adler2017solving,Jin2017,wang2020deep, arridge2019solving}. A common idea is to use a neural network as a post-processing step to improve the quality of the CT reconstruction \cite{pelt2018improving}. While this approach can produce promising results in reducing noise or imaging artefacts, such post-processing neural networks are generally considered ``black-box''. Additionally, such networks typically require large amounts of training data, which can be impractical in certain applications, for instance in medical imaging. Also, the post-processing solution might not be optimal for severely ill-posed problems, where the quality of the initial reconstruction is extremely flawed. 

An alternative approach is to combine analytical inverse problems algorithms with deep learning. These kinds of reconstruction techniques take advantage of the strengths of both variational regularization and deep learning. Having the explicit forward operator improves robustness and generalizability of the network, compared to previously mentioned ``black-box'' neural networks. In addition, such model-based reconstruction reduces the amount of training data needed. This is due to the fact that the network has less parameters, and the forward operator encodes the imaging geometry so it does not need to be learned \cite{adler2017solving, adler2018learned}. 

With modern machine learning methods, it has become possible to suppress streak artefacts with unprecedented efficiency.
As shown in \cite{bubba2019learning}, it is possible to learn only the information which is not stably represented in the limited-angle tomographic data. The authors propose a hybrid framework with shearlets and a convolutional neural network for solving limited-angle CT problems. There, the network fills in only the missing shearlet coefficients, and the rest is done using rigorous inversion methods. Thus, the result can be seen as interpretable, as the network only affects a controlled part of the final reconstruction. In \cite{adler2018learned}, the authors propose a deep learning framework for CT by unrolling a proximal-dual optimization method. In their Learned Primal-Dual (LPD) algorithm, convolutional neural networks replace the proximal operators. In \cite{andrade2021deep}, the authors combine microlocal analysis to the LPD algorithm and propose a scheme for joint reconstruction and wavefront set inpainting, enabling significant reduction of streak artefacts.

Inherent in deep learning is the aforementioned ``black box'' problem; in other words, poor understanding of how did the algorithm arrive at the given reconstruction. Especially in medical imaging, it is often important to use algorithms that are as interpretable as possible. With this is mind, our present approach aims to only learn the unstable parts of the boundaries, which then can be displayed as extra information, on top of a classical (non-learned) reconstruction. 


\begin{figure}
\begin{picture}(320,150) 
\put(50,0){\includegraphics[height=6cm]{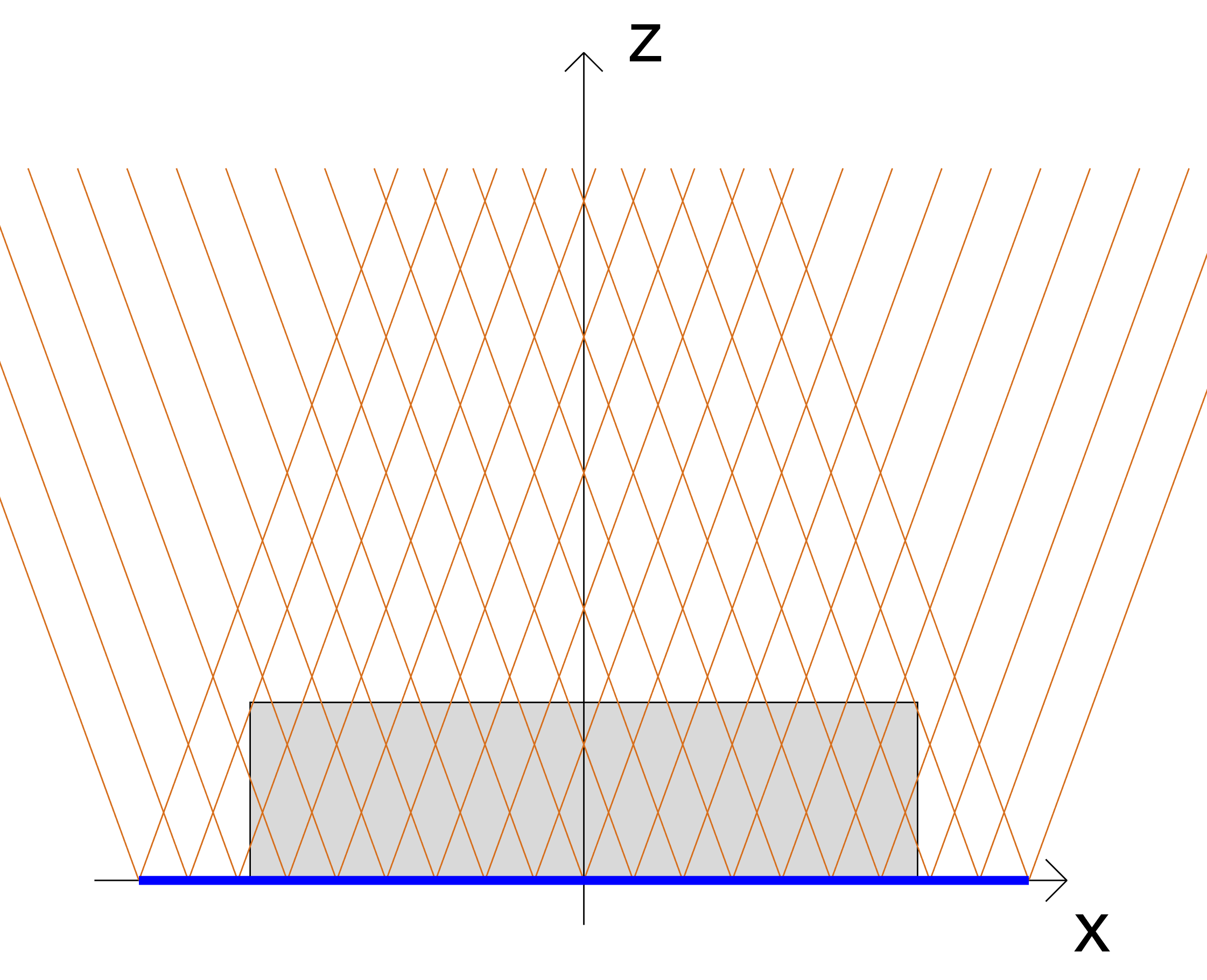}}
\end{picture}    
    \caption{
    Idealized limited-angle imaging geometry we consider in our computational examples. We restrict to parallel-beam geometry. Shown here are the two extreme angles in our measurement model; we use a total of 50 directions distributed evenly in the angular variable between these two extremes. There is a significant difference in terms of imaging technology between our model and the two applications shown in Figure \ref{fig:DBTgeom}, due to the fan-beam geometry used there. Namely, it is cumbersome to realize parallel-beam measurements with an X-ray source. However, the mathematics of wavefront sets is essentially the same, so our findings should apply with straightforward modifications.  }
    \label{fig:our_geom}
\end{figure}

We introduce a new algorithm for improved reconstruction: a data-driven and geometry-guided hybrid. We first compute an initial reconstruction of the target using a suitable algorithm (in our case Primal-Dual Fixed-Point (PDFP) method \cite{chen2013primal} promoting frame sparsity). Due to the limited-angle imaging geometry, the reconstruction contains stretching artefacts. 

Next, we aim to detect the stable parts of boundaries within the reconstruction. We transform the reconstruction using Dual-Tree Complex Wavelet Transform (DTCWT), which has directional selectivity oriented at $\pm 15^{\circ}$, $\pm 45^{\circ}$ and $\pm 75^{\circ}$. DTCWT does not offer such a theoretically complete discretisation of the wavefront (WF) set as curvelets and shearlets do. However, we combine DTCWT with nonlinear morphological operations and a neural network model in a novel way, resulting in a robust six-angle edge detector, capable of automatically excluding streak artefacts. Our learned edge detection algorithm is independently applicable to a wide range of image processing tasks. 


Using a neural network model, we can then learn the missing parts of the wavefront set in the framework of six oriented subbands, extending it beyond the scope of the limited-angle measurements. Our approach manages to ignore the streak artefacts and estimates the unstable parts of the boundaries of objects with unprecedented efficiency.
Then, we can use this extended information about the wavefront set to form a boundary estimate that can be added as an overlay to the PDFP reconstruction. 


We illustrate our method with computational examples, working with parallel-beam geometry as shown in Figure \ref{fig:our_geom}. This  simplifying assumption helps in computational experiments as we can divide the 3D tomography problem defined in $xyz$-space into a stack of independent 2D problems posed in $xz$-planes determined by a constant value of $y$. The results are viewed by browsing through the $xy$-planes of the reconstructed volume by varying the $z$-coordinate. Our algorithm leads to remarkable accuracy in separating objects located away from each other in the $z$-coordinate but having overlapping projections in the $xy$-plane. Such objects are typically merged together in traditional limited-angle reconstructions. We note that there is no obstruction in principle in extending our methodology to fan-beam or cone-beam geometries.  


The structure of this paper is as follows. In section \ref{section2}, we go through the main mathematical background information necessary to understand the proposed method. These topics include dual-tree complex wavelet transform (section \ref{section:DTCWT}), wavefront set, singular support and tomography (section \ref{section:wst}), Primal-Dual Fixed-Point optimization (section \ref{section:PDFP}), and finally, we explain the morphological operations used in the computations (section \ref{section:morphological_operations}). 
In section \ref{learnedWFsection}, we describe our proposed method for learned wavefront set extraction for limited-angle tomography. Then, in section \ref{Sec:nonmicro}, we continue to explain the process of learning a microlocal prior. 
Next, in section \ref{section:results}, we show the results of our proposed method in the $xz$-plane (section \ref{section:results2d}) and in the $xz$-plane (section \ref{section:results3d}) for a simplistic three-dimensional computational phantom. Finally, in section \ref{section:discussion}, we end with some concluding remarks.

\clearpage
\section{Mathematical background} \label{section2}

\subsection{Dual-tree complex wavelet transform} \label{section:DTCWT}



\begin{figure}[b!]
\begin{picture}(400,200)
\put(0,20){\includegraphics[width=0.33\textwidth]{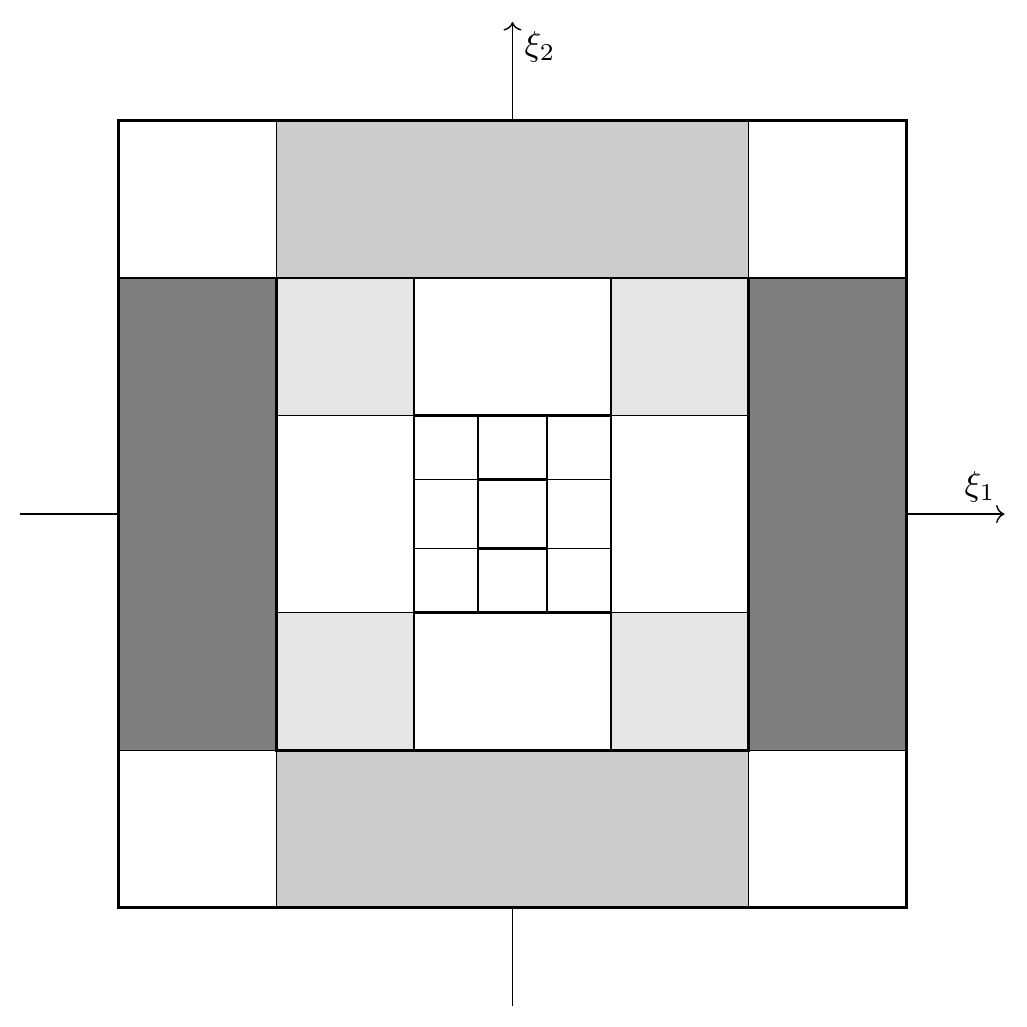}}
\put(160,20){\includegraphics[width=0.33\textwidth]{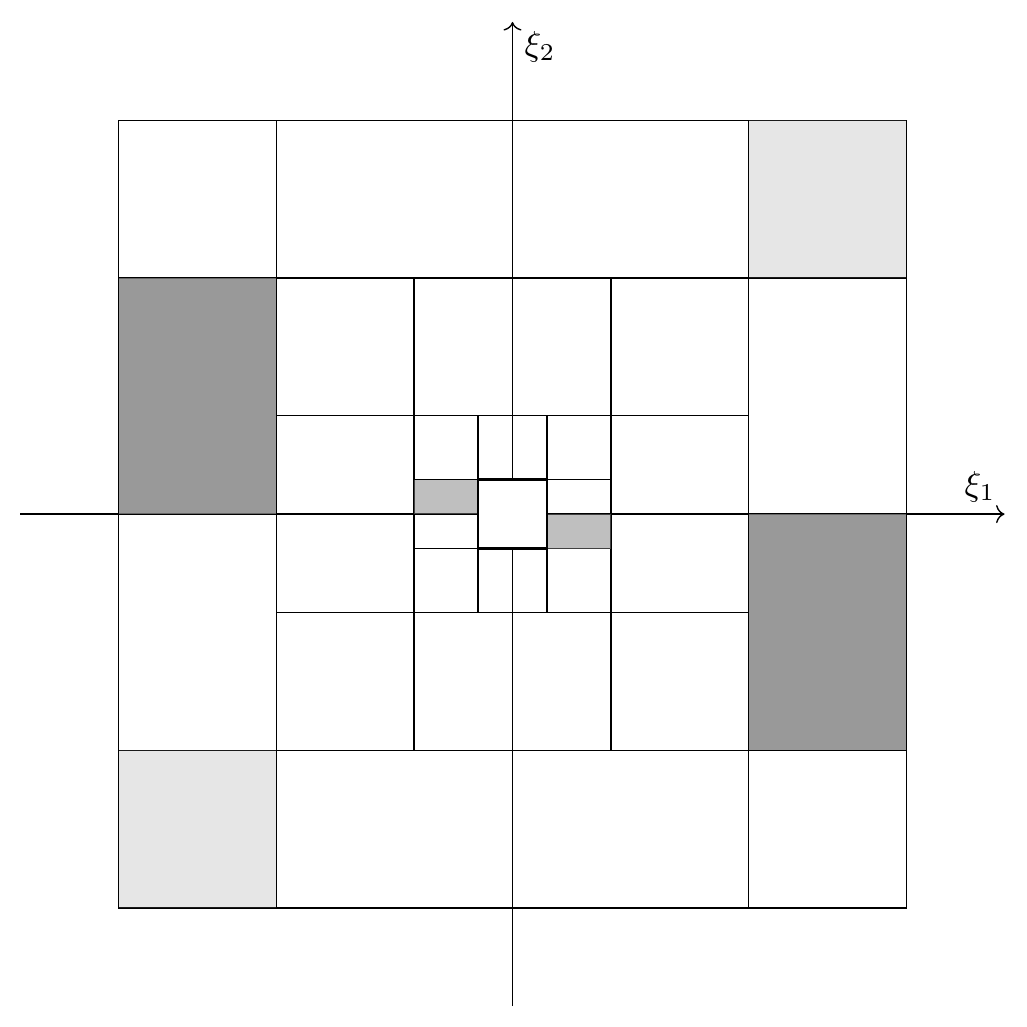}}
\put(320,20){\includegraphics[width=0.33\textwidth]{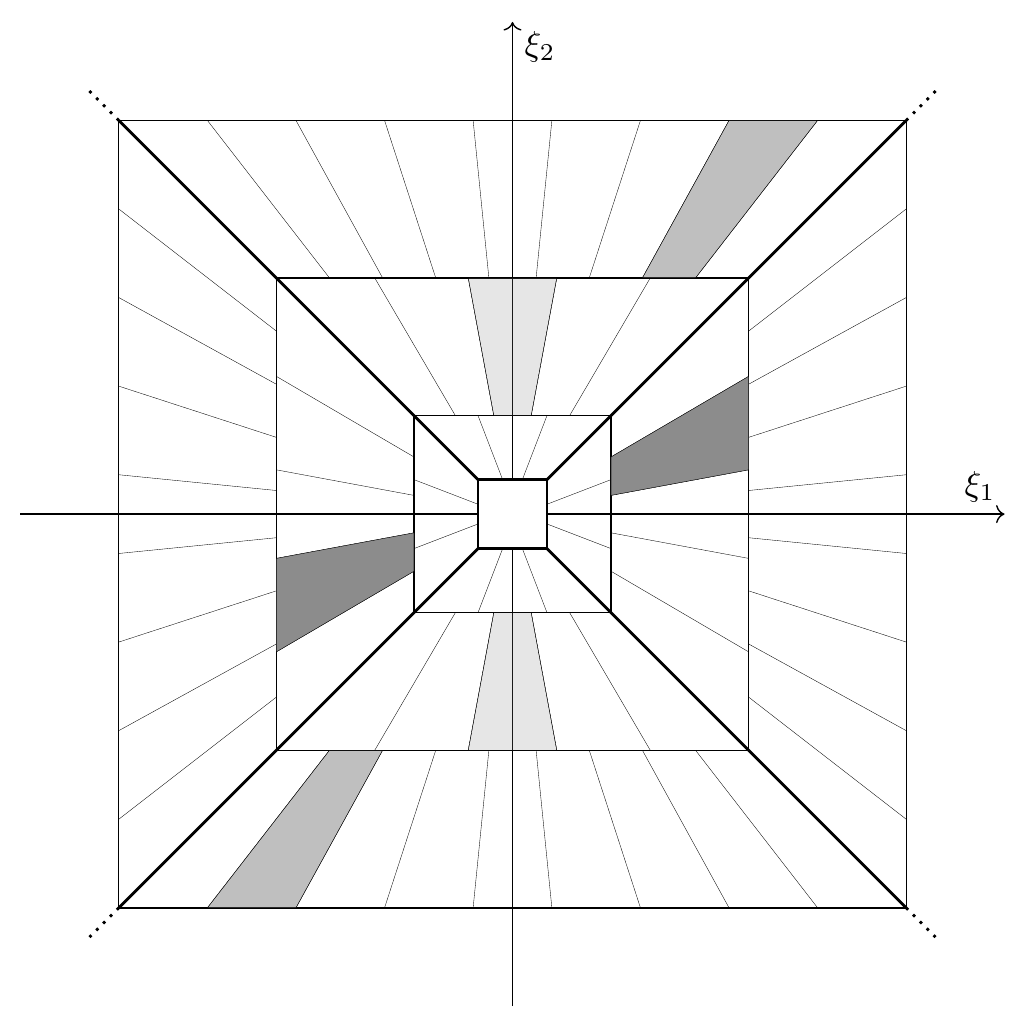}} 
\put(20,0){(a) Real-valued wavelets }
\put(72,140){HL} 
\put(72,45){HL} 
\put(22,95){LH} 
\put(115,95){LH} 
\put(22,140){HH}
\put(115,140){HH}
\put(22,45){HH}
\put(115,45){HH}
\put(170,0){(b) Complex-valued wavelets}
\put(180,110){$\overline{H}L$}
\put(180,140){$\overline{H}H$}
\put(212,140){$\overline{L}H$}
\put(212,45){$LH$}
\put(245,140){$LH$}
\put(245,45){$\overline{L}H$}
\put(275,140){$HH$}
\put(275,45){$\overline{H}H$}
\put(180,45){$HH$}
\put(180,75){$HL$}
\put(275,75){$\overline{H}L$}
\put(275,110){$HL$}
\put(340,0){(c) Shearlets}
\end{picture}
\caption{Frequency tilings of different multi-resolution transforms. (a) Real-valued wavelets are often understood to give horizontal, vertical, and diagonal details. This can be seen in the approximate supports of the basis functions in the frequency domain. (b) The approximate supports of the Fourier transforms of the frame functions corresponding to the six subbands in the dual-tree complex wavelet transform. (c) Shearlets offer very detailed analysis of orientations using parabolic scaling and increasing numbers of orientations with finer scales. However, the benefits come with a computational cost.}
\label{fig:cw_ferqs}
\end{figure}
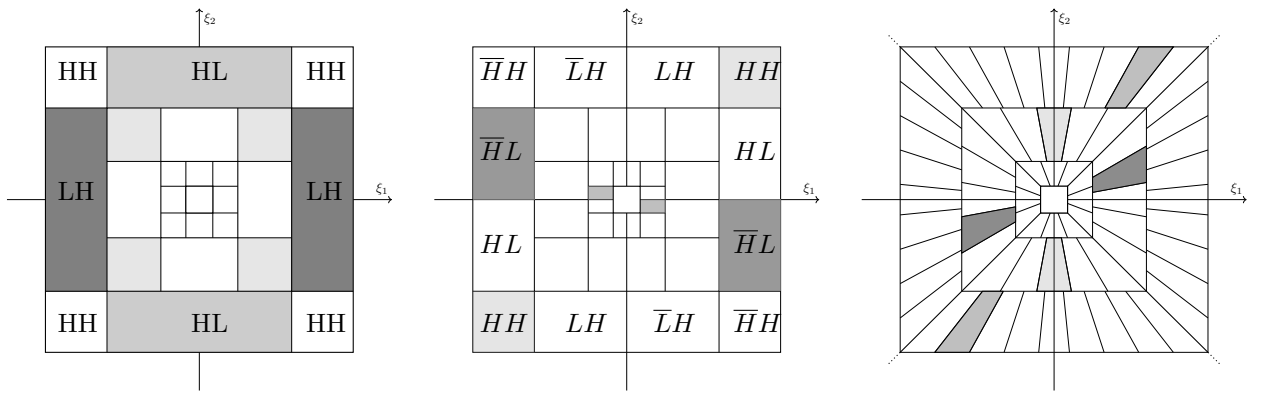


The dual-tree complex wavelet transform consists of a complex-valued scaling function $\phi$ and a complex-valued wavelet $\psi$. 
Consider a 
complex and analytic wavelet $\psi(x)$, given by  
$$\psi(x)  = \psi_{h}(x) + i \psi_{g}(x)$$
and associated with a high-pass filter $H$, and a complex and analytic scaling function $\phi(x)$, that is given by
$$\phi(x)  = \phi_{h}(x) + i \phi_{g}(x)$$
and associated with a low-pass filter $L$. 
Note that $\psi_{h}$ and $\phi_{h}$ are the real parts of $\psi(x)$ and $\phi(x)$, and $\psi_{g}$ and $\phi_{g}$ are the imaginary parts of $\psi(x)$ and $\phi(x)$, respectively. For computational purposes, we use finitely supported wavelets and thus approximately analytic wavelets.

We denote the complex wavelet coefficients by $d_\nu (j,n)\in\C$, where $j=0,\dots,J$ represents the number of scales used in the complex wavelet representation and
\begin{equation}\label{oriented_subbands}
\nu\in \mathcal{I}=\{\overline{L}H, \overline{H}H, \overline{H}L, HL, HH ,LH\}
\end{equation}
represents the index of the six oriented subbands in each scale $j$. Note that $j=0$ corresponds to the mother wavelet and $j = J$ corresponds to the finest scale. See Figure \ref{fig:cw_ferqs}b for the frequency-domain content indexed by $\nu$. 

To build intuition, the expression for the 2D wavelet in subband $HH$ is obtained by:
\begin{eqnarray}
    \psi(x_1,x_2) &=& [\psi_h(x_1) +i \psi_g(x_2)][\psi_h(x_1) +i \psi_g(x_2)] \\
    &=& \psi_h(x_1) \psi_h(x_2) - \psi_g(x_1) \psi_g(x_2) + i[\psi_g(x_1) \psi_h(x_2) + \psi_h(x_1) \psi_g(x_2)].
\end{eqnarray}
Note that this wavelet is oriented at $-45^{\circ}$. To obtain the 2D wavelet oriented at $+45^{\circ}$, that is the subband $\overline{H} H$, we compute $\psi_2(x_1,x_2) = \overline{\psi(x_1)} \psi(x_2)$, where
\begin{eqnarray}
    \psi_2(x_1, x_2) &=& [\overline{ \psi_h(x_1) + i \psi_g(x_1)}][\psi_h(x_2) + i \psi_g(x_2)] \\
    &=& \psi_h(x_1) \psi_h(x_2) + \psi_g(x_1) \psi_g(x_2) + i[\psi_h(x_1) \psi_g(x_2) - \psi_g(x_1) \psi_h(x_2)].
\end{eqnarray}
In a similar manner, the six subbands of the 2D complex wavelets for each scale $j$ are computed as follows:
\begin{eqnarray}
\label{subbandLH}
 \Psi_{LH,j,n}(x_1,x_2) &:=& \phi(2^j x_1-n_1) \psi(2^j x_2-n_2) \\
\label{subbandHL}
  \Psi_{HL,j,n}(x_1,x_2) &:=& \psi(2^j x_1-n_1) \phi(2^j x_2-n_2) \\
\label{subbandHH}
  \Psi_{HH,j,n}(x_1,x_2) &:=& \psi(2^j x_1-n_1) \psi(2^j x_2-n_2) \\
\label{subbandHHbar}
  \Psi_{\overline{H} H,j,n}(x_1,x_2) &:=& \overline{\psi(2^j x_1-n_1)} \psi(2^j x_2-n_2) \\
\label{subbandLHbar}
    \Psi_{\overline{L}H,j,n}(x_1,x_2) &:=& \overline{\phi(2^j x_1-n_1)} \psi(2^j x_2-n_2) \\
\label{subbandHLbar}
  \Psi_{\overline{H}L,j,n}(x_1,x_2) &:=& \overline{\psi(2^j x_1-n_1)} \phi(2^j x_2-n_2) .
\end{eqnarray}
Here $n=(n_1,n_2)\in\Z^2$. For discrete images of the size of power-of-two (which we consider throughout the paper), $n_1$ and $n_2$ range from $0$ to $2^j-1$, depending on the scale $j$. In our computations, j will range from $0$ to $6$, $0$ corresponding to the coarsest scale, and $6$ corresponding to the finest scale. We remark that the digital wavelet transform starts by computing the finest scale by applying filters to a pixel image. The calculation progresses towards coarser scales by repeatedly applying the filters to the low-pass subbands. For a more detailed look on the theoretical properties of the complex wavelet transform, please refer to \cite{selesnick2005dual}.

\begin{figure}[t]
\centering
\begin{subfigure}{.15\textwidth}
  \centering
  \includegraphics[width=23mm]{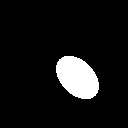}
  \caption{Phantom}
  \label{fig:cw_ph}
\end{subfigure}%
\begin{subfigure}{.15\textwidth}
  \centering
  \includegraphics[width=23mm]{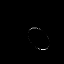}
  \caption{$\overline{L}H$}
  \label{fig:cw1}
\end{subfigure}%
\begin{subfigure}{.15\textwidth}
  \centering
  \includegraphics[width=23mm]{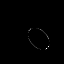}
  \caption{$\overline{H}H$}
  \label{fig:cw2}
\end{subfigure}%
\begin{subfigure}{.15\textwidth}
  \centering
  \includegraphics[width=23mm]{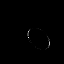}
  \caption{$\overline{H}L$}
  \label{fig:cw3}
\end{subfigure}%
\begin{subfigure}{.15\textwidth}
  \centering
  \includegraphics[width=23mm]{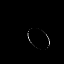}
  \caption{$HL$}
  \label{fig:cw4}
\end{subfigure}%
\begin{subfigure}{.15\textwidth}
  \centering
  \includegraphics[width=23mm]{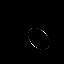}
  \caption{$HH$}
  \label{fig:cw5}
\end{subfigure}%
\begin{subfigure}{.15\textwidth}
  \centering
  \includegraphics[width=23mm]{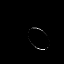}
  \caption{$LH$}
  \label{fig:cw6}
\end{subfigure}%
\caption{The first scale of complex wavelet detail coefficients of a phantom shown in (a). The orientations of coefficient subbands are denoted below each subband. Here, the absolute value of the complex-valued detail coefficients is shown.}
\label{fig:cw_detail_coeffs}
\end{figure}

The complex wavelet coefficients are defined by:
\begin{equation}
d_\nu(j,n):=\langle f,\Psi_{\nu,j,n}\rangle =
\int_{-\infty}^\infty\int_{-\infty}^\infty f(x_1,x_2) \Psi_{\nu,j,n}(x_1,x_2)\,dx_1dx_2.
\end{equation}
Here, the function $f$ can be recovered from the coefficients by
\begin{equation}
f(x_1,x_2) = \sum_{\nu\in J}\sum_{j=1}^J\sum_{n_1=0}^{2^j-1}\sum_{n_2=0}^{2^j-1} d_\nu(j,n) S^{-1}\Psi_{\nu,j,n}(x_1,x_2),
\end{equation}
where $S$ is the frame operator
$$
 S = T^\ast T,
$$
where $T$ is the analysis operator and $T^{\ast}$ is the synthesis operator corresponding to $S$.
See Figure \ref{fig:cw_detail_coeffs} for an example of the absolute values of complex wavelet detail coefficients for all six subbands of the finest scale.

The complex wavelet transform also gives rise to real-valued, final-level low-pass scaling coefficients, known as the approximation coefficients. They are given by:

\begin{equation}
    c(n)  = \int_{-\infty}^\infty\int_{-\infty}^\infty f(x_1,x_2) \phi(x_1-n_1) \phi(x_2-n_2)\,dx_1dx_2,\quad n=(n_1,n_2). 
\end{equation}






The motivation for using complex wavelets is that they offer a good compromise between directionality and computational cost. More precisely, real-valued wavelets offer only very limited information about the orientation of the edges in an image. On the other hand, curvelets \cite{candes2004new} and shearlets \cite{kutyniok2012shearlets,labate2005sparse} use parabolic scaling, along with rotation or shearing, to achieve remarkably accurate approximation properties for the wavefront set. However, for our purposes, the DTCW transform offers just the right balance between directionality and computational cost. See Figure \ref{fig:cw_ferqs} for an illustration of the frequency content of the building blocks of these three types of transforms. 

\subsection{Wavefront set, singular support and tomography} \label{section:wst}

Microlocal analysis tells us that limited-angle tomography data specifies certain elements of the wavefront set of the X-ray attenuation coefficient, while information about the rest of the wavefront set is not present 
in any well-posed form. We recall briefly the definitions of singular support and wavefront set here for the convenience of the reader. 

The wavefront set of a signal $f$ describes both the location $x_0$ and the direction $\theta_0$ of singularities. The signal $f$ is smooth near $x_0$ if there is a cutoff function $\phi \in C^{\infty}_{c}$, $\phi(x_0) \neq 0$, such that the Fourier transform of $\phi f$ decays rapidly. That is,
$$ \widehat{\phi f}(\xi) = \mathcal{O} (|\xi|)^{-N} \quad |\xi| \rightarrow \infty \, \text{for all} \, N >0.$$
The signal $f$ has a singularity in $x_0$ if for all cutoff functions $\phi$, the Fourier transform of $\phi f$ does not decay rapidly. The set of all singularities of $f$ is called the \textit{singular support} and is denoted by $\texttt{singsupp}(f)$. To define the orientation of the singularities $x \in \texttt{singsupp} (f)$, we look for directions along which the localized Fourier transform $\widehat{\phi f}$ does not decay rapidly. 


Define the wavefront set of a function $f$ as the set $WF(f)(x_0, \alpha_0)$ with location $x_0 \in \texttt{singsupp} (f)$ and direction $\alpha_0$, such that for all cutoff functions $\phi \in C^{\infty}_c$, $\phi (x_0) \neq 0$ , the localized Fourier transform $\widehat{\phi f}(\xi)$ does not decay rapidly in any polar wedge $W_{\delta} = \{(r,w):|w -\alpha_0| < \delta \}$, where $(r,w)$ are the polar coordinates in the frequency domain. The direction $\alpha_0$ of a singularity $x_0 \in \texttt{singsupp} {f}$ can be considered as the direction of maximum change of $f$ at $x_0$. In particular, for a piecewise constant function having a jump along a Jordan curve $\gamma$, $\texttt{singsupp} (f)$ equals $\gamma$ and $WF(f)$ consists of pairs $(x_0, N(x_0))$ with $x_0\in\gamma$ and $N(x_0)$ the normal vector of $\gamma$ at $x_0$.

Consider the continuous two-dimensional Radon transform $\mathcal{R}f$ of a function $f$:
$$
  \mathcal{R} f (\theta,s) = \int_{L(\theta,s)} f(x) dS(x), 
$$ 
where $L(\theta,s) = \{ x \in \mathbb{R}^2 : x_1 \text{cos}(\theta) + x_2 \text{sin}(\theta) = s \}$ is the line with normal direction $\theta$ and signed distance from the orientation $s$. Given the Radon transform data $\mathcal{R} f(\theta,s)$ for $(\theta,s)$ arbitrarily near $(\theta_0, s_0)$, one can reconstruct singularities of $f$ stably with location $x \in L(\theta_0,s_0)$ and direction $\theta_0$. In other words, a singularity $(x_0,\theta_0)$ of $f$ is visible from a limited angle measurement if the line $L(\theta_0, x_0\cdot \theta_0)$ is recorded in the measurements. See \cite{greenleaf1989nonlocal,quinto1993singularities,frikel2013characterization,frikel2013sparse} for references. 

\begin{figure}
\begin{picture}(320,100) 
\put(-10,0){\includegraphics[width=12.5cm]{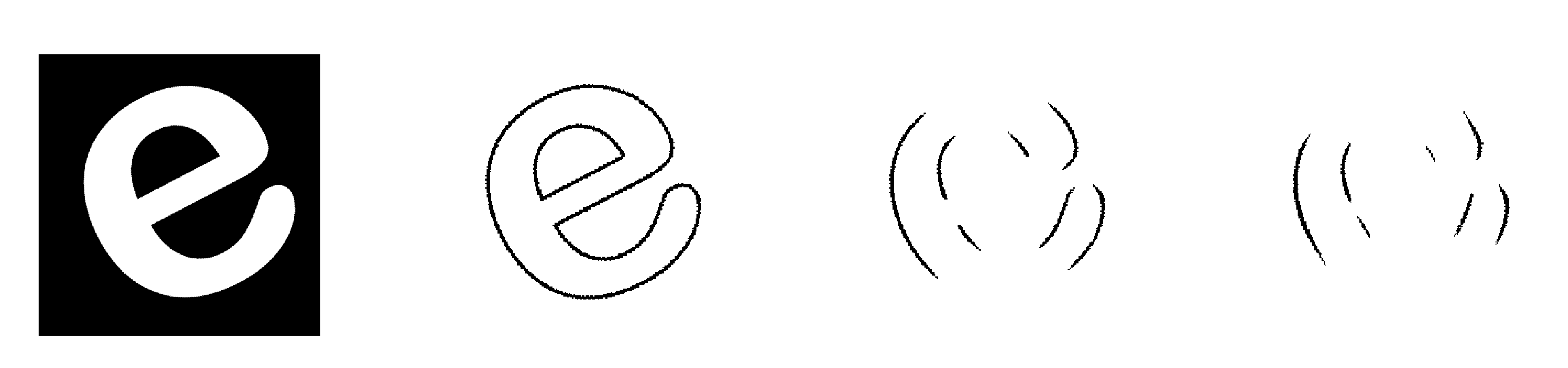}}
\put(0,90){(a) Phantom}
\put(91.7,90){(b) Full angle}
\put(183.3,90){(c) $|\theta|\leq 45^\circ$}
\put(275,90){(d) $|\theta|\leq 20^\circ$}
\end{picture}
\caption{Demonstration of the visible part of the wavefront set in limited-angle tomography. (a) Digital phantom, where attenuation coefficient $f(x)=0$ in the black area and $f(x)=1$ inside the white character ``e''. (b) Full singular support of $f$. (c) We simulated parallel-beam Radon transform data of $f$ in the limited angular range $-45^\circ\leq\theta\leq 45^\circ$, calculated the filtered back-projection reconstruction, and determined the singular support using dual-tree complex wavelets. Which parts of the jump curves are missing? Those whose tangents are not parallel to any X-rays in the data. (d) Same as (c) but with more limited angular range $-20^\circ\leq\theta\leq 20^\circ$.}
\label{fig:visibleWF}
\end{figure}
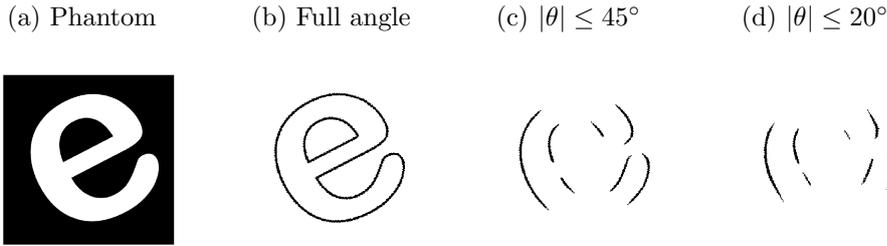

In (parallel-beam) limited-angle tomography, we know the Radon transform $\mathcal{R} f (\theta,s)$ for all $s\in\R$ and for the angular range $[\theta_0-\delta,\theta_0+\delta]$, for some $\theta_0\in\R$ and $0<\delta<\pi$. Then, the visible part of the wavefront set is $WF(f)(x_0, \alpha)$, for 
$$
\pi/2+\theta_0-\delta \leq \alpha \leq \pi/2+\theta_0+\delta.
$$
In Figure \ref{fig:visibleWF}, we show an example of the visible part of the wavefront set for $\delta=\pi/4$, (45°), and $\delta=\pi/18$, (20°). 


\subsection{Primal-dual fixed-point (PDFP) optimization} \label{section:PDFP}


In this study, we use the Primal-dual fixed-point algorithm, described by Chen, Huang and Zhang \cite{chen2013primal}, to solve the following optimization problem. In the discrete setting, the inverse problem of reconstructing a tomographic image ${\bf{f}} \in \mathbb{R}^n$ based on X-ray measurements ${\bf{m}} \in \mathbb{R}^{m}$ is modeled by 
\begin{equation}\label{discrete_inv_prob}
    {\bf{m}}  = A {\bf{f}}+ \epsilon,
\end{equation}
where $A \in \mathbb{R}^{m \times n}$ is a discretized linear forward operator and $\epsilon>0$ models additive Gaussian noise. We consider regularized solutions to the inverse problems, achieved by minimizing the following functional:
\begin{equation} \label{min_problem}
 {\bf {f}}_S =   \bigg\{ \argmin_{{\bf{f}} \in \mathbb{R}^{n}, {\bf{f}} \geq 0} \frac{1}{2} \| A{\bf{f}} - m\|_2^2 + \mu \| W_\mathbb{C}{\bf {f}}\|_{1} \bigg \},
 \end{equation}
where the term $\| W_\mathbb{C}{\bf{f}}\|_{1}$ promotes solutions having sparse representation in the wavelet basis, and $\mu>0$ serves as a regularization parameter providing a balance between data fidelity and \emph{a priori} information. The component-wise inequality ${\bf{f}} \geq 0 $  is based on the physical fact that X-radiation can only attenuate inside the target and not strengthen. Analogous regularization was used in \cite{purisha2017controlled}, but with real wavelets instead of complex.



The PDFP algorithm can be used to iteratively solve the above minimization problem (equation \ref{min_problem}). The algorithm is given by:
\begin{equation}\label{PDFP-alg}
 \left\{ \begin{array}{ll}
{\bf{y}}^{k+1} &= \mathbb{P}_C ({\bf{f}}^k-\tau \bigtriangledown {\mathcal{G}}({\bf{f}}^{(k)})-\lambda (W_\mathbb{C})^T {\bf{v}}^k),\\
{\bf{v}}^{k+1} &= (I - \mathcal{T}_{\mu\tau / \lambda}) (W_\mathbb{C} {\bf{y}}^{k+1}+{\bf{v}}^k),\\
{\bf{f}}^{k+1} &= \mathbb{P}_{C} ({\bf{f}}^k-\tau \bigtriangledown {\mathcal{G}}({\bf{f}}^{(k)})-\lambda (W_\mathbb{C})^T {\bf{v}}^{(k+1)}),
\end{array} \right.
\end{equation}
where $\tau$ and $\lambda$ are positive parameters, ${\mathcal{G}}({\bf{f}}) = \frac{1}{2} \|A{\bf{f}} -{\bf{m}}\|_2^2$, and $\mu > 0 $ represents the regularization parameter. Parameters 
$0<\lambda< \frac{1}{\lambda_{\text{max}}(W_\mathbb{C}(W_\mathbb{C})^T)}$,
where $\lambda_{\text{max}}$ is the maximum eigenvalue, and 
$0<\tau <2/\tau_{\text{lip}}$, 
where  $\tau_{\text{lip}}$ is the Lipschitz constant for $\mathcal{G}({\bf{f}})$, need to be suitably chosen for convergence. The soft-thresholding operator $\mathcal{T}$ is defined radially for complex-valued inputs as
\begin{equation}
\mathcal{T}_{\mu}(\theta) = (|\theta| - \mu)e^{i\theta}. 
\end{equation}
In equation \eqref{PDFP-alg}, the non-negative quadrant is denoted by $C = \mathbb{R}^{N^2}_{+}$ and $\mathbb{P}_C$ is the Euclidean projection, that is, the operator $\mathbb{P}_C$ replaces any non-negative elements in the input vector by zeroes.

\subsection{Morphological operations} \label{section:morphological_operations}

In this work, we use morphological operations \cite{gonzalez2018digital} to clean and pre-process greyscale and binary data. In mathematical morphology, we consider the original image as the object $\mathcal{D}\subset \mathbb{Z}^2$. A structuring element $S\subset \mathbb{Z}^2$ is a pre-defined binary image that is used to probe the object $\mathcal{D}$, and conclude how it fits the shapes of the object. 

The morphological dilation operation can be used to "grow" objects in a binary image. The shape and size of the structuring element $S$ defines the extent of the dilation. The dilation operation is defined as
\begin{equation} \label{dilation}
    \mathcal{D} \oplus S = \bigcup_{s\in S} \mathcal{D}_s,
\end{equation}
where $\mathcal{D}_s = \{ d+s | d \in \mathcal{D}, s \in S\}$ is the translation of $\mathcal{D}$ by the structuring element $s$.

On the other hand, morphological erosion thins out objects in a binary image. It is the dual operation of dilation. The erosion operation is defined as
\begin{equation} \label{erosion}
    \mathcal{D} \ominus S = \bigcap_{s\in S} \mathcal{D}_{-s}.
\end{equation}

The morphological opening operation removes small details from the object but preserves the shape and size of larger objects. The opening of $\mathcal{D}$ by $S$ is defined as 
\begin{equation} \label{opening}
    \mathcal{D} \circ S = (\mathcal{D} \ominus S) \oplus S,
\end{equation}
where $\mathcal{D} \ominus S$ is the morphological erosion of $\mathcal{D}$ by $S$, and $\oplus$ is the dilation operation of the result by $S$. The opening of $\mathcal{D}$ by $S$ can be seen as the union of all the translations of $S$ so that it fits within the object $\mathcal{D}$:
\begin{equation}\label{open}
    \mathcal{D} \circ S = \bigcup_{z\in \mathcal E_{\mathcal D}}S_z,
\end{equation} 
where $ \mathcal E_{\mathcal D}=\{z\in \mathbb Z^2:\ S_z \subseteq \mathcal{D}\}$.

The morphological skeleton operation extracts the centerline of the objects in a binary image while preserving its topology. The operation transforms the object $\mathcal{D}$ to 1-pixel wide curved lines, while not changing the essential structure. It can be expressed in terms of morphological erosions and openings:
\begin{equation}\label{morph_skeleton}
    Skeleton(\mathcal{D}) = \bigcup_{k=0}^K (\mathcal{D} \ominus kS) - (\mathcal{D} \ominus kS) \circ S,
\end{equation}
where $(\mathcal{D} \ominus kS)$ expresses $k$ successive erosion operations and $K= \max \{k | (\mathcal{D} \ominus kS) \neq \emptyset \}$ denotes the last step before the object $\mathcal{D}$ erodes to an empty set.

Morphological operations can be extended for greyscale images. In that case, $\mathcal{D}$ in formula \eqref{open} is a greyscale pixel image. In greyscale morphology, images $d$ are functions mapping the Euclidean space $E$ into $\mathbb{R} \cup \{-\infty,\infty\}$. That is, $ d: E \rightarrow \mathbb{R} \cup \{-\infty,\infty\}. $ The greyscale structuring elements are also functions mapping Euclidean space into $\mathbb{R} \cup \{-\infty,\infty\}$. 

Denote an image by $d(x)$ and a structuring function by $b(x)$. Then, the dilation operation is defined as
\begin{equation}\label{gray_dilation}
    (d \oplus b)(x)  = \sup_{y \in E} [d(y) + b(x-y)],
\end{equation}
where $\sup$ denotes the supremum. The erosion of $d$ by $b$ is given by
\begin{equation}\label{gray_erosion}
    (d \ominus b)(x)  = \inf_{y \in E} [d(y) - b(x-y)],
\end{equation}
while the opening and closing are given by
\begin{equation}\label{grey_opening}
    d \circ b  = (d \ominus b) \oplus b,
\end{equation}
\begin{equation}\label{grey_closing}
    d \circ b  = (d \oplus b) \ominus b.
\end{equation}

If $b(x)={\bf 1}_B(x)$ is the indicator function of the set $B$
and $d(x)={\bf 1}_D(x)$ is the indicator function of the set $D$,
we have 
\begin{equation}\label{indicator functions}
    (d \circ b) (x)  = {\bf 1}_{D \circ B}(x).
\end{equation}
Similar formulas hold for dilation $d \oplus b$ and erosion $d \ominus b$.

\clearpage

\section{Learned wavefront set extraction} \label{learnedWFsection}

\begin{figure}\label{PDFP_reco}
\centering
\begin{subfigure}{.33\textwidth}
  \centering
  \includegraphics[height=46mm]{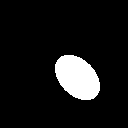}
  \caption{ }
  \label{fig:gt_phantom}
\end{subfigure}%
\begin{subfigure}{.33\textwidth}
  \centering
  \includegraphics[height=46mm]{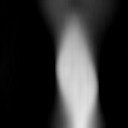}
  \caption{ }
  \label{fig:initial_pdfp}
\end{subfigure}
\caption{(a) Elliptical digital phantom with constant attenuation and zero background. (b) PDFP reconstruction with complex wavelet regularization from 50 X-ray projection images over a 40-degree opening angle. The central direction of projection is vertical. Note the vertical elongation or stretching of the ellipse in the reconstruction. This is a typical artefact in limited-angle tomography.}
\label{fig:phantom_and_reco}
\end{figure}

In limited-angle tomography, the reconstructed image typically fails to reproduce certain parts of the wavefront set of the original target, as explained in section \ref{section:wst}. This is often accompanied by severe stretching artefacts in the reconstruction along the central direction of projection, as illustrated in Figure \ref{fig:phantom_and_reco}. 

\begin{figure}[t]
    \centering
    \includegraphics[width=8cm]{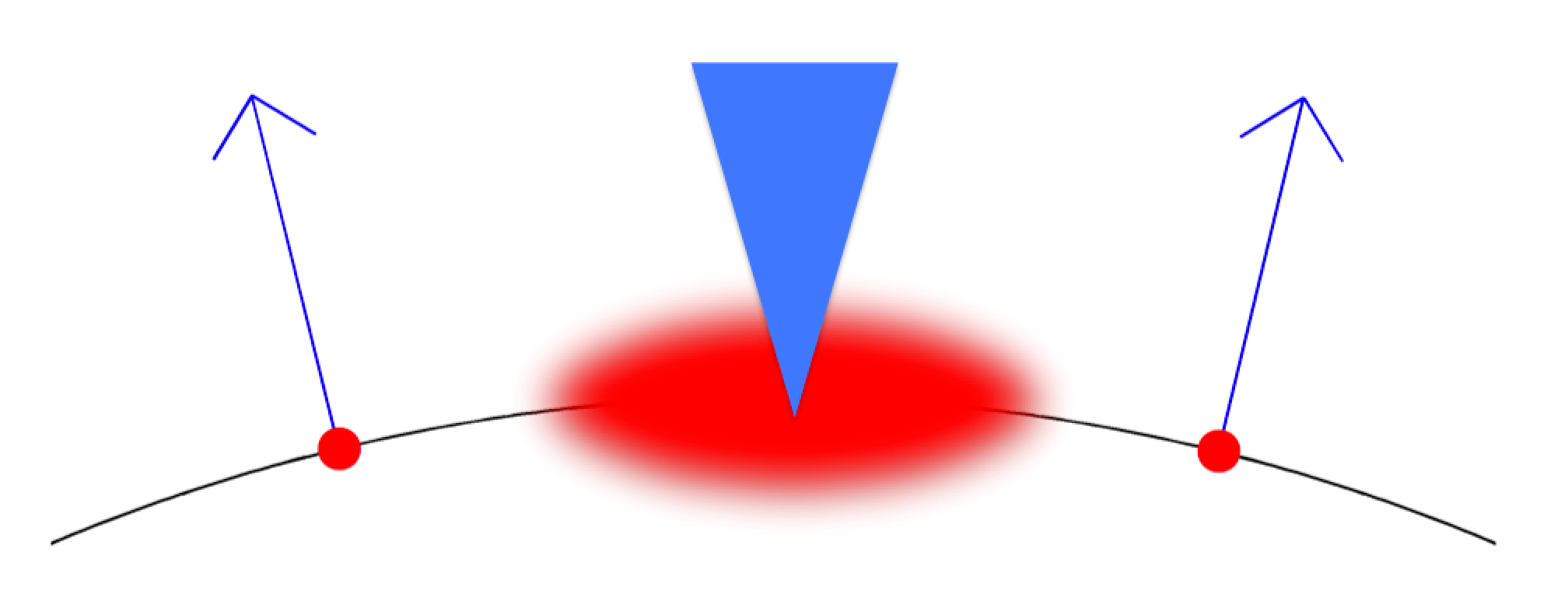}
    \caption{The geometric idea behind our microlocal prior. Assume we know two elements of the wavefront set, located spatially at the two red dots and having the directions indicated by the two blue arrows. We work with piece-wise constant attenuation coefficients having reasonably regular boundary curves (shown as the black arc here) between the constant domains. Then we know that there should be another element in the wavefront set with a spatial location somewhere in the are shown as blurred red, and with direction within the range indicated with the blue triangle.}
    \label{fig:microlocalprior}
\end{figure}

Our goal is to fill in the missing parts of the wavefront set by machine-learning the geometric rules of how the known parts of the wavefront set extend into the unknown parts. See Figure \ref{fig:microlocalprior} for the geometric idea behind this microlocal prior. After we have the full wavefront set available, we can use it to form a boundary estimate of features.

We compute the aforementioned boundary estimate in several steps, including the application of two distinct convolutional neural networks. (a) The first neural network learns to extract the known part of the wavefront set by transforming morphologically opened complex wavelet coefficients to a binary form. (b) The second neural network learns to predict the complete wavefront set from the incomplete wavefront set that has been morphologically dilated.
We will discuss the second network, among with the operations related to it, in Section \ref{Sec:nonmicro}.


\subsection{Complex wavelet coefficients of a limited-angle reconstruction}
We simulate a parallel-beam limited-angle sinogram with 50 projection directions and a 40-degree opening angle. Further, we compute a reconstruction using variational optimization with complex wavelet regularization, as described in section \ref{section:PDFP}. 
Because of the limited-angle imaging geometry, the features in the reconstruction are stretched along the central direction of projection, as can be seen in Figure \ref{fig:phantom_and_reco}.

\begin{figure}
\centering
\begin{subfigure}{\textwidth}
  \centering
  \includegraphics[width=165mm]{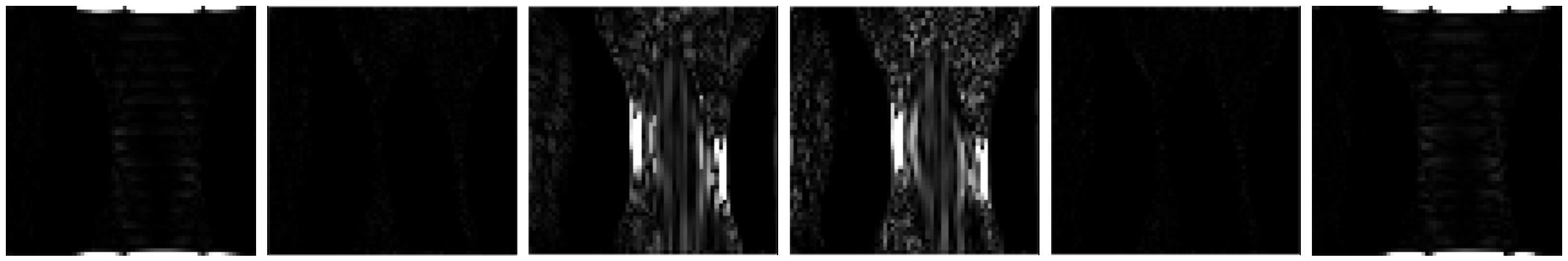}
  \caption{Absolute values of the detail coefficients $A_{\nu,6}(n_1,n_2) = 
|{d}_{\nu} (6,n_1,n_2)|$ of a limited-angle PDFP reconstruction.}
  \label{fig:abs_coeffs}
\end{subfigure}%
\newline
\vspace{0.5cm}
\begin{subfigure}{\textwidth}
  \centering
  \includegraphics[width=165mm]{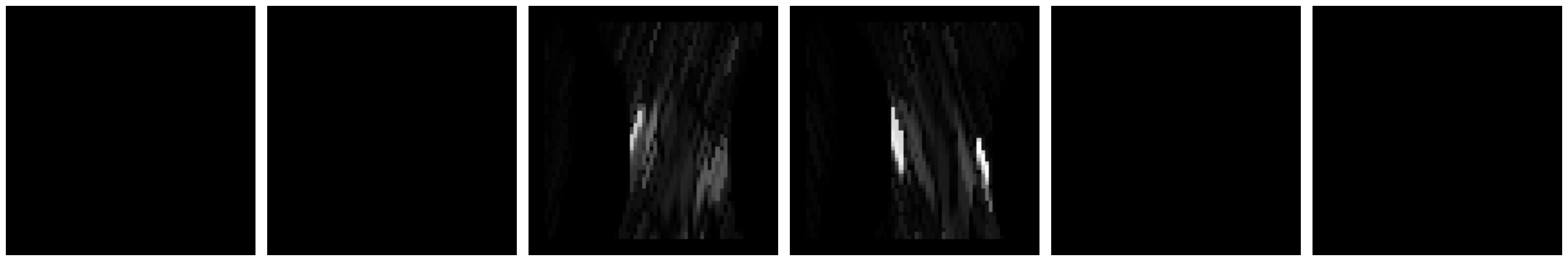}
  \caption{Cleaned coefficients $\widetilde{A}_{\nu,6}$ obtained by morphological opening with a line structuring element $S_{\nu,6}$. Input for the neural network $\mathcal{N}_1$.}
  \label{fig:cleaned_coeffs}
\end{subfigure}%
\newline
\vspace{0.5cm}
\begin{subfigure}{\textwidth}
  \centering
  \includegraphics[width=165mm]{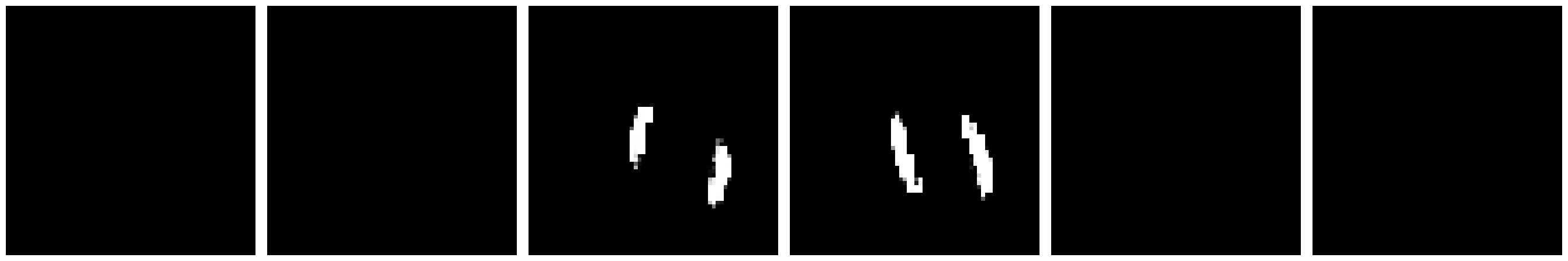}
  \caption{Binary mask $\widehat{A}_{\nu,6}(n_1,n_2)$, output from neural network $\mathcal{N}_1$ that performs binary thresholding.}
  \label{fig:nn1_output1}
\end{subfigure}
\newline
\caption{Going from (a) the imperfect and approximate wavefront set provided by DTCW coefficients into a robust wavefront set estimator. The technique involves two nonlinear steps: (b) morphology followed by (c) learning. The subband indexes from left to right are: $\overline{L}H, \overline{H}H, \overline{H}L, HL, HH ,LH$.}
\end{figure}

First, we want to extract the wavefront set (see Section \ref{section:wst}) from the reconstruction. This is accomplished by picking out a discrete approximation of the wavefront set by using the finest scale of dual-tree complex wavelet coefficients $d_\mathbb{\nu}(j,n_1,n_2)$, where $j = 6$, $n=(n_1,n_2)\in \Z^2$, and $\nu \in \mathcal I$ is as described in section \ref{section:DTCWT}. 
Then, we take the absolute value of the coefficients: 
\begin{equation}\label{abscomplexwav}
A_{\nu,6}(n_1,n_2) = 
|{d}_{\nu} (6,n_1,n_2)|.
\end{equation} 
A demonstration can be seen in Figure \ref{fig:abs_coeffs}. 

Due to the limited-angle data, we observe nonzero complex wavelet coefficients mainly in subbands $d_{\overline{H}L}(6,n_1,n_2)$ and $d_{HL}(6,n_1,n_2)$, while subbands $d_{\overline{L}H}(6,n_1,n_2)$, $d_{\overline{H}H}(6,n_1,n_2)$, $d_{HH}(6,n_1,n_2)$, and $d_{LH}(6,n_1,n_2)$ contain no information beyond noise. This is because the reconstruction from a limited-angle sinogram can only encode information about certain directions of the wavefront set, as explained in Section  \ref{section:wst}. 






\subsection{Morphological opening of the complex wavelet coefficients} 

\begin{figure}[t]
\centering
\begin{subfigure}{.15\textwidth}
  \centering
  \includegraphics[height=23mm]{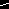}
  \caption{$\overline{L}H$}
\end{subfigure}%
\begin{subfigure}{.15\textwidth}
  \centering
  \includegraphics[height=23mm]{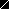}
  \caption{$\overline{H}H$}
\end{subfigure}%
\begin{subfigure}{.15\textwidth}
  \centering
  \includegraphics[height=23mm]{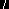}
  \caption{$\overline{H}L$}
\end{subfigure}%
\begin{subfigure}{.15\textwidth}
  \centering
  \includegraphics[height=23mm]{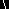}
  \caption{$HL$}
\end{subfigure}%
\begin{subfigure}{.15\textwidth}
  \centering
  \includegraphics[height=23mm]{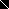}
  \caption{$HH$}
\end{subfigure}%
\begin{subfigure}{.15\textwidth}
  \centering
  \includegraphics[height=23mm]{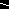}
  \caption{$LH$}
\end{subfigure}
\caption{\label{Fig:line_St_elems}Line structuring elements $S_{\nu,6}$ that are binary images, for each index $\nu$ of the oriented subbands, listed in (\ref{oriented_subbands}). 
See (\ref{subbandLH})-(\ref{subbandHLbar}) for the connection of the orientations of the line structuring elements to specific wavelet functions. }
\label{fig:line_SE}
\end{figure}

Due to stretching artefacts and imperfect measurements, the complex wavelet coefficients available in subbands $d_{\overline{H}L}(6,n_1,n_2)$ and $d_{HL}(6,n_1,n_2)$ are noisy. Therefore, we improve this information of the wavefront set using
two nonlinear steps: grayscale morphology followed by learning.

To extract useful information about the complex wavelet coefficients, we use the morphological opening operation on the absolute values of the complex wavelet coefficients.
First, we set subbands $d_{\overline{L}H}(6,n_1,n_2), d_{\overline{H}H}(6,n_1,n_2), d_{HH}(6,n_1,n_2)$, and $d_{LH}(6,n_1,n_2)$ to zero. 
For the remaining two subbands $d_{\overline{H}L}(6,n_1,n_2)$ and $d_{HL}(6,n_1,n_2)$, we perform morphological opening with oriented line structuring elements $S_{\nu,6}$: 

\begin{equation}\label{Morp_open}
\widetilde{A}_{\nu,6} = A_{\nu,6} \circ S_{\nu,6},
\end{equation}
where $A_{\nu,6}$, defined in equation (\ref{abscomplexwav}), is a grayscale image considered as a function $A_{\nu,6}:\Z^2\to [0,1]$.  The line structuring elements $S_{\nu,6}$ are the binary images shown in Figure \ref{Fig:line_St_elems} and are considered as a function $S_{\nu,6}:\Z^2\to \{0,1\}$.
For our limited-angle problem, we use structuring elements $S_{\overline{H}L,6}$ and $S_{HL,6}$, since we perform opening for subbands $d_{\overline{H}L}(6,n_1,n_2)$ and $d_{HL}(6,n_1,n_2)$. The result is illustrated in Figure \ref{fig:cleaned_coeffs}.

Now, one would think that hard-thresholding of the opened subbands is as follows:
\begin{equation*}
\left\{ \begin{array}{ll}
1, & \widetilde{A}_{\nu,6} (n_1,n_2) \geq \epsilon,\\
0 & \textrm{otherwise,}
\end{array} \right.
\end{equation*} 
would give binary indicators for the discrete wavefront set. This would be the case if we had a good threshold $\epsilon>0$ available. However, in practice, imperfect knowledge of the noise amplitude and the absence of ground truth makes it difficult to design an automatic method for choosing the threshold. Therefore, we resort to learning the thresholding process.

\subsection{Learning the binary mask} \label{sec:learning_the_binary_mask}



We use a convolutional neural network $\mathcal{N}_1:\R^{\Z^2\times \mathcal I}\to [0,1]^{\Z^2\times \mathcal I}$
to learn the binary indicators for the discrete wavefront set
in the two subbands $d_{\overline{H}L}(6,n_1,n_2)$ and $d_{HL}(6,n_1,n_2)$. The input of neural network  $\mathcal{N}_1$ is the tensor $\widetilde A=(\widetilde A_{\nu,6}(n))_{n\in \Z^2,\nu \in \mathcal I}$ obtained from equation (\ref{Morp_open}) and the network maps it to $\widehat A=(\widehat A_{\nu,6}(n))_{n\in \Z^2,\nu \in \mathcal I}$, that is,
$$
\widehat A=\mathcal{N}_1(\widetilde A).
$$
The training of the neural network uses the morphologically opened complex wavelet coefficients $\widetilde{A}_{\overline{H}L,6}(n_1,n_2)$ and $\widetilde{A}_{HL,6}(n_1,n_2)$.
After the network $\mathcal{N}_1$ performs the thresholding, it outputs a binary mask 
with ones indicating the location of the wavefront set. See Figure \ref{fig:nn1_output1} for an example result.




Training inputs are generated as follows. We first generate 5000 two-dimensional phantoms of size $128 \times 128$ with ellipse shapes that have varying radius, tilt, and position. For each phantom, we consider 50 X-ray measurements over a $40$-degree opening angle, spanning from $70$ to $110$ degrees. 
We add $5 \%$ noise to the measurement sinogram. Then, we compute the reconstruction of each phantom using the PDFP algorithm with complex wavelet regularization. Next, we compute the absolute values of the complex wavelet coefficients, $A_{\nu,6}(n_1,n_2)$, of each reconstruction.
We set subbands $A_{\overline{L}H,6}(n_1,n_2), A_{\overline{H}H,6}(n_1,n_2), A_{HH,6}(n_1,n_2)$, and $A_{LH,6}(n_1,n_2)$ to zero. We perform morphological opening of the subbands $A_{\overline{H}L,6}(n_1,n_2)$ and $A_{HL,6}(n_2,n_2)$ by using formula \eqref{Morp_open}. 
These morphologically opened sets of subbands $\widetilde{A}_{\nu,6}$ serve as the training inputs.


The corresponding ground truth values 
are generated as follows. First, we compute the absolute value of the complex wavelet coefficients of the ground truth phantoms. We again set subbands $d_{\overline{L}H}(6,n_1,n_2), d_{\overline{H}H}(6,n_1,n_2), d_{HH}(6,n_1,n_2)$, and $d_{LH}(1,n_1,n_2)$ to zero. We perform morphological opening on subbands $d_{\overline{H}L}(6,n_1,n_2)$ and $d_{HL}(6,n_1,n_2)$ using \eqref{Morp_open} to get ${\widetilde{A}}_{\overline{H}L,6} (n_1,n_2)$ and ${\widetilde{A}}_{HL,6} (n_1,n_2)$. Then, we convert the data into a binary form by thresholding sufficiently large values of the morphologically opened coefficients to $1$ and the rest are set to $0$. That is, in training of neural network $\mathcal{N}_1$, the ground truth data is obtained using the formula
\begin{equation} \label{eq:gt_threshold}
\overline{A}_{\nu,6}(n_1,n_2) = \left\{ \begin{array}{ll}
1, & \widetilde{A}_{\nu,6} (n_1,n_2) \geq \epsilon,\\
0 & \textrm{otherwise}
\end{array} \right.
\end{equation} 
using a suitably chosen $\epsilon>0$.

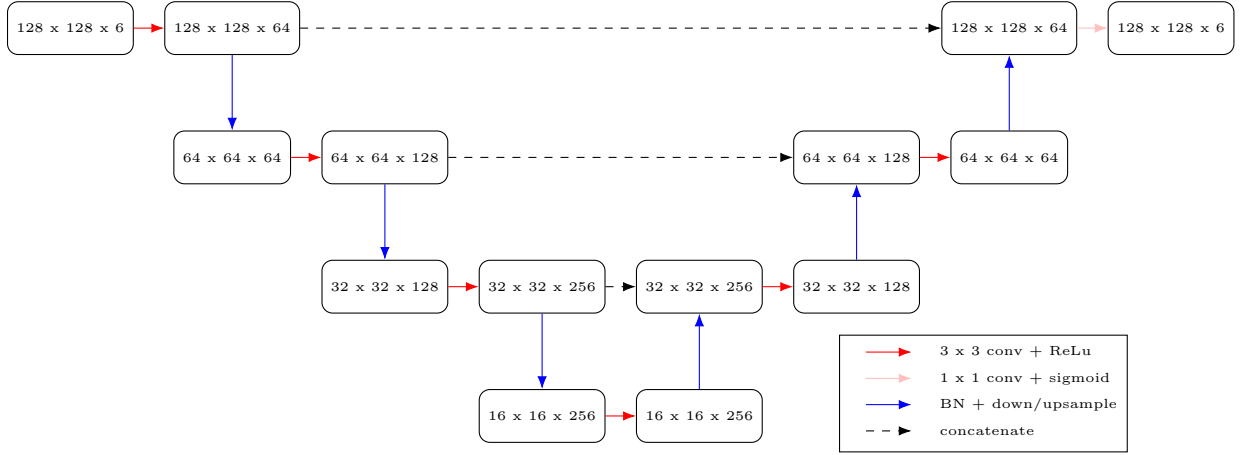
\begin{figure*}
\input{architecture1}
\caption{Network architecture of $\mathcal{N}_1$.}
\label{fig:nn1_arch}
\end{figure*}




We implement neural network $\mathcal{N}_1$ as a convolutional autoencoder with residual connections, as inspired by \cite{ronneberger2015u}. For the encoder and decoder parts of the network, we use convolution kernels of size $3 \times 3$, stride of size 1, and rectified linear unit (ReLU) activation functions. After each convolutional layer, we perform batch normalization, followed by down- or up-sampling. We use skip connections to join the encoder and decoder parts of the network to improve training. For the last layer, we use a $1 \times 1$ convolution with a sigmoid activation. See the detailed architecture in Figure \ref{fig:nn1_arch}. The best results were achieved by using the dice coefficient loss function given by:

\begin{equation}
    L(\overline{A},\widehat{A}) = 1 - \frac{2\sum(\overline{A} \cdot \widehat{A})}{\sum(\overline{A} + \widehat{A})}, 
\end{equation}
where $\overline{A}$ denotes the ground truth values and $\widehat{A}$ the network predictions. 

The training input and output datasets were of size $5000 \times 128 \times 128 \times 6$ each. During training, 500 training set examples were used as the validation set.

\begin{figure}[b!]
    \centering
    \includegraphics[width=\textwidth]{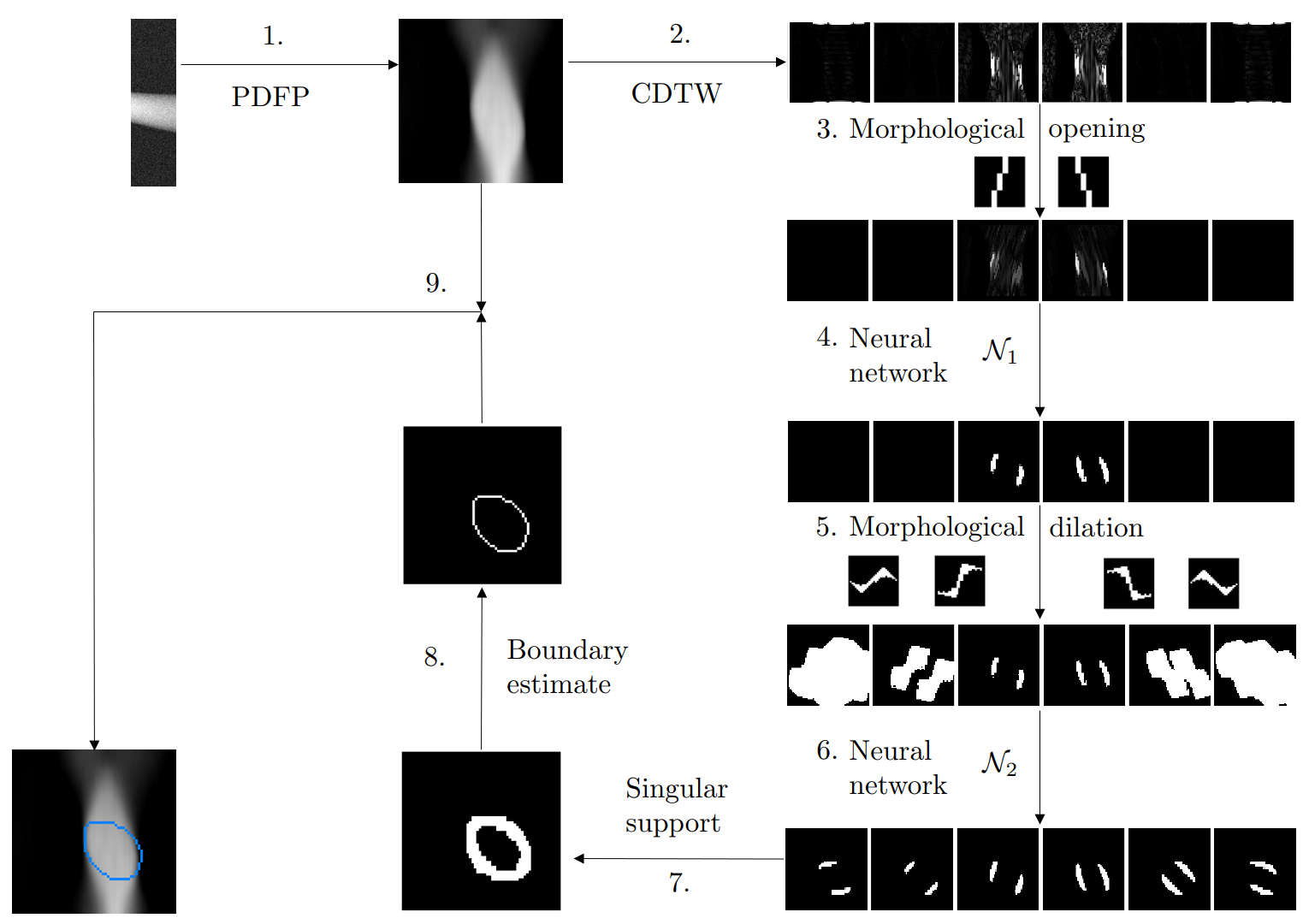}
    \caption{Workflow of the proposed method: 1. Compute an initial reconstruction using the PDFP algorithm with complex wavelet regularization. 2. Compute the finest scale complex wavelet coefficients for the six subbands and take their absolute value. 3. Clean the coefficients using morphological opening with appropriately oriented line structuring elements. 4. Give this to the first neural network, which thresholds the six subbands into a binary format. 5. Compute an initial guess of the microlocal prior, by dilating the binary subbands with specific directed structuring elements. 6. Give this initial guess for the second neural network, which outputs a prediction of the wavefront set in all six subbands. 7. Combine the predicted wavefront set information in all six subbands to form the singular support. 8. Compute the morphological skeleton of the singular support, estimating its boundary. 9. Add this learned boundary estimate as an overlay on top of the PDFP reconstruction.}
    \label{fig:flowchart}
\end{figure}

\begin{figure}[b!]
\centering
\begin{subfigure}{.24\textwidth}
  \centering
  \includegraphics[height=34mm]{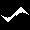}
  \caption{$\widetilde{S}_{\overline{H}H\rightarrow \overline{L}H}$}
\end{subfigure}%
\begin{subfigure}{.24\textwidth}
  \centering
  \includegraphics[height=34mm]{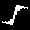}
  \caption{$\widetilde{S}_{\overline{H}L \rightarrow \overline{H}H}$}
\end{subfigure}%
\begin{subfigure}{.24\textwidth}
  \centering
  \includegraphics[height=34mm]{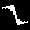}
  \caption{$\widetilde{S}_{HL \rightarrow HH}$}
\end{subfigure}%
\begin{subfigure}{.24\textwidth}
  \centering
  \includegraphics[height=34mm]{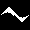}
  \caption{$\widetilde{S}_{HH \rightarrow LH}$}
\end{subfigure}
\caption{The custom-made directional structuring elements $\widetilde{S}_{\nu\rightarrow\nu^\prime}$ we use for our specific limited-angle measurement geometry.}
\label{fig:SE}
\end{figure}

\clearpage
\section{Learned microlocal prior} \label{Sec:nonmicro}

In the previous section, we explained how to extract the known part of the wavefront set from a limited-angle tomography reconstruction. In this section, we will go through the rest of the steps in order to learn a boundary estimate. See Figure \ref{fig:flowchart} for a summary of the entire workflow of our proposed method.

The basic idea behind our microlocal prior is based on the assumption of the smoothness and bounded curvature of the singular support of the X-ray attenuation coefficient. When a part of a wavefront set is detected in a subband, it is approximately a piece of straight line, following the singular support along a direction designated for that subband. Imagine traveling along the line. Eventually the singular support is bound to turn outside the orientation range of the subband you are on, as the interfaces between areas of different attenuation are closed curves. Then the line ends abruptly. But there is a continuation of it on one of the two neighboring subbands, depending on the direction of the turn. The representation of the singular support continues there as an almost straight line (along an orientation 30 degrees rotated from where you started), until another inevitable turn is forcing a jump to yet another subband. The structuring elements shown in Figure \ref{fig:SE} are designed so as to capture the locations in a neighboring subband where the continuation can go.

We learn the boundary estimate with a second neural network $\mathcal{N}_2$, using the output of the first neural network $\mathcal{N}_1$. In order to do so, we first approximate the microlocal prior by morphological dilation. 
The details of the steps are explained in the sections below.


\begin{figure}[b!]
\centering
\begin{subfigure}{\textwidth}
  \centering
  \includegraphics[width=165mm]{images/output_nn1_new.png}
  \caption{Binary mask $\widehat{A}_{\nu,6}(n_1,n_2)$, starting point for the dilation operation. Output of the first neural network $\mathcal{N}_{1}$.}
  \label{fig:nn1_output}
\end{subfigure}
\newline
\vspace{0.5cm}
\begin{subfigure}{\textwidth}
  \centering
  \includegraphics[width=165mm]{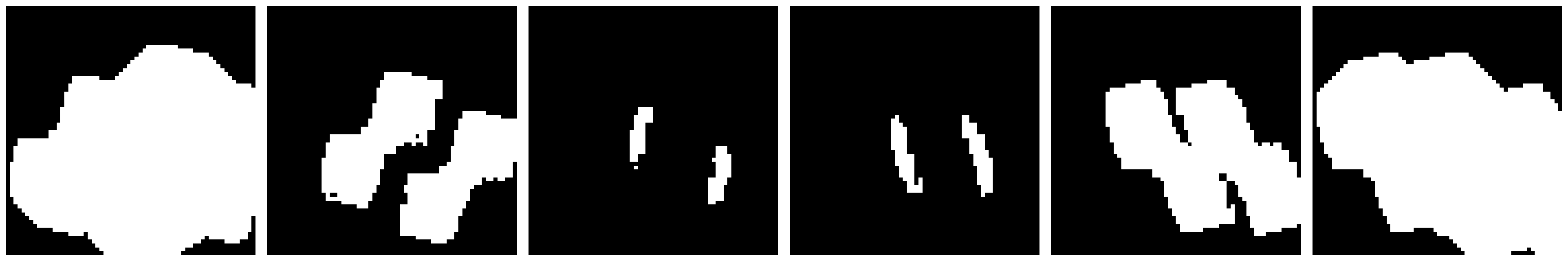}
  \caption{Initial guess of the wavefront set $D_{\nu,6}(n_1,n_2)$, computed using morphological dilation. Input for the second network, $\mathcal{N}_{2}$.}
  \label{fig:nn1_output_predictions}
\end{subfigure}%
\newline
\vspace{0.5cm}
\begin{subfigure}{\textwidth}
  \centering
  \includegraphics[width=165mm]{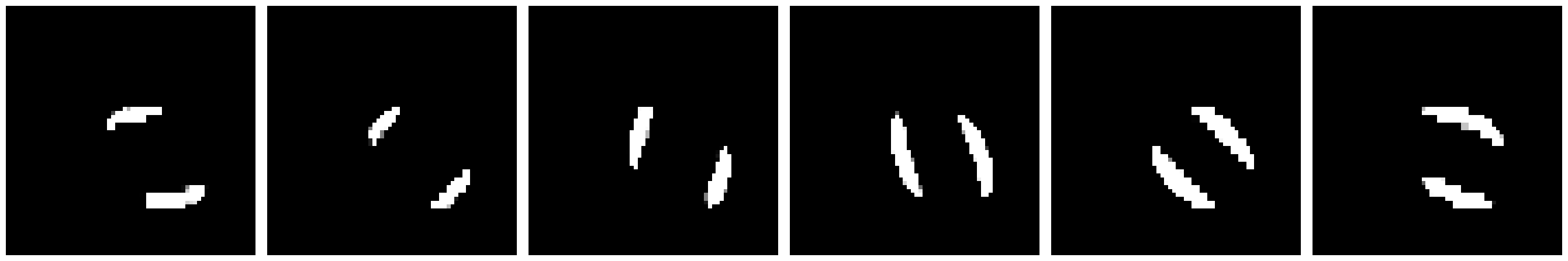}
  \caption{Full prediction of the wavefront set $\widehat{D}_{\nu,6}(n_1,n_2)$, output of the second network $\mathcal{N}_2$.}
  \label{fig:nn2_output}
\end{subfigure}%
\newline
\caption{Learning to fill in the incomplete wavefront set in six complex wavelet coefficient subbands, extending it to the entire domain. The subband indexes from left to right are: $\overline{L}H, \overline{H}H, \overline{H}L, HL, HH ,LH$.}
\end{figure}

\subsection{Approximation of the microlocal prior by morphological dilation} \label{sec:dilation}

Now, we extend the wavefront set information from the two available subbands (see Figure \ref{fig:nn1_output}) to the remaining four subbands. 
We begin by using morphological dilation, as explained in section \ref{section:morphological_operations}, to compute an initial guess for the location of nonzero coefficients. As the structuring elements for the operation, we have designed custom-made directional structuring elements that are able to estimate the location of the wavefront set based on a neighbouring subband. The structuring elements $\widetilde{S}_{\nu\rightarrow\nu^\prime}$, $\nu,\nu^\prime\in \mathcal I$ that we use for our specific limited-angle measurement geometry are shown in Figure \ref{fig:SE}. These structuring elements are designed so that the elements (a) and (b) in Figure \ref{fig:SE} are the rotated and scaled images of the set 
$$S^-=\{(x,y)\in \R^2:\ x\in [-1,1], y=-a\,\hbox{sign}(x)x^2 \hbox{ with some }0\le a\le 1\} $$ 
and the elements (c) and (d) are the rotated and scaled images of the set  
$$S^+=\{(x,y)\in \R^2:\ x\in [-1,1],\ y=+a\,\hbox{sign}(x)x^2 \hbox{ with some }0\le a\le 1\}.$$ 

The design is motivated by the assumption that the unknown object may have a discontinuity on a curve $\gamma$ whose curvature is bounded. For example, the unknown object corresponding to function $f$ may be a sum of the indicator function of a smooth domain $D$ multiplied by a smooth function $f_1$ and a smooth function $f_2$:
$$ f(x_1,x_2)=f_1(x_1,x_2){\bf  1}_D(x_1,x_2)+f_2(x_1,x_2),$$
and $\gamma:[-1,1]\to \R^2$ is a path satisfying $\gamma([-1,1])\subset \partial  D$. 
Let us consider the case when the path $\gamma$ is parametrized as $\gamma(s)=(s,h(s))$, $s\in [-1,1]$, where $h:[-1,1]\to \R$ is a function that satisfies $h(0)=0$ and $h'(0)=0$. Moreover, assume that the normal vector of $\gamma(x)$ turns clockwise when  $s>0$ and $s$ grows. Then for $s>0$ the second derivative of $h$ satisfies $ h''(s)<0$. If we also have $0> h''(s)\ge -1$, then $\gamma([0,1])\subset S^-$. When $s<0$,  we assume that the normal vector of $\gamma(x)$ turns clockwise when $-s$ grows and that $0<h''(s)\le 1$, in which case  $\gamma([-1,0])\subset S^-$. Note that for example the structuring element $\widetilde S_{\overline H L\to \overline H H}$ corresponds to predicting the points $x=(x_1,x_2)$ on the discontinuity $\partial  D$ with a normal vector $\nu(x)$ of $\partial  D$ such that there exists a nearby point $\tilde x=(\tilde x_1,\tilde x_2)\in \partial  D$ with a normal vector $\nu(\tilde x)$ such that $\nu(x)$ is a vector close to $\nu(\tilde x)$ that has been turned in the clockwise direction (see Figure \ref{fig:microlocalprior}).

The approximation of the microlocal prior is done as follows using the above structuring elements. We make no changes to the subbands $\overline{H}L$ and $HL$, since we have already extracted the wavefront set for these subbands in the previous steps. That is,
\begin{eqnarray}
D_{\overline{H}L,6} (n_1,n_2) &=& \widehat{A}_{\overline{H}L,6} (n_1,n_2) \\
D_{HL,6} (n_1,n_2), &=& \widehat{A}_{HL,6} (n_1,n_2).
\end{eqnarray}
To get the wavefront set estimate for the adjacent subbands $\overline{H}H$ and $HH$, we dilate subbands $\overline{H}L$ and $HL$ with the structuring elements $\widetilde{S}_{\overline{H}L \rightarrow \overline{H}H}$ and $\widetilde{S}_{HL \rightarrow HH}$, respectively. That is, 
\begin{eqnarray}
D_{\overline{H}H,6} &=& D_{\overline{H}L,6} \oplus \widetilde{S}_{\overline{H}L \rightarrow \overline{H}H} \\
D_{HH,6} &=& D_{HL,6} \oplus \widetilde{S}_{HL \rightarrow HH}.
\end{eqnarray} 
Next, for subbands $\overline{L}H$ and $LH$, we treat the once dilated subbands $\overline{H}H$ and $HH$ as the objects for the dilation with structuring elements $\widetilde{S}_{\overline{H}H \rightarrow \overline{L}H}$ and $\widetilde{S}_{HH \rightarrow LH}$, respectively. That is, 
\begin{eqnarray}
D_{\overline{L}H,6} &=& D_{\overline{H}H,6} \oplus \widetilde{S}_{\overline{H}H \rightarrow \overline{L}H}  
= \big(D_{\overline{H}L,6} \oplus \widetilde{S}_{\overline{H}L \rightarrow \overline{H}H} \big) \oplus \widetilde{S}_{\overline{H}H \rightarrow \overline{L}H} \\
D_{LH,6} &=& D_{HH,6} \oplus \widetilde{S}_{HH \rightarrow LH}  
= \big(D_{HL,6} \oplus \widetilde{S}_{HL \rightarrow HH} \big) \oplus \widetilde{S}_{HH \rightarrow LH}. 
\end{eqnarray} 

Now, we have a coarse initial guess of the wavefront set for the network $\mathcal{N}_2$ to refine. See Figure \ref{fig:nn1_output_predictions} for an example.

\subsection{Learning to extend the wavefront set into the entire domain}

We want to learn to refine the wavefront set estimate with neural network $\mathcal{N}_2$, based on the approximation provided by the dilation operation. The network $\mathcal{N}_2$ outputs a prediction of the full wavefront set $\widehat{D}_{\nu,6}(n_1,n_2)$, extending it to all six subbands. We use a convolutional neural network $\mathcal{N}_2:\R^{\Z^2\times \mathcal I}\to [0,1]^{\Z^2\times \mathcal I}$, where the input of $\mathcal{N}_2$ is the tensor $D=(D_{\nu,6}(n))_{n\in \Z^2,\nu \in \mathcal I}$ and the network maps it to $\widehat D=(\widehat D_{\nu,6}(n))_{n\in \Z^2,\nu \in \mathcal I}$. That is,
$$
\widehat D=\mathcal{N}_2( D).
$$
See Figure \ref{fig:nn2_output} for an example output.

To get the training inputs for network $\mathcal{N}_2$, we use the same phantoms as with network $\mathcal{N}_1$. We perform morphological dilation (as explained in section \ref{sec:dilation}) on thresholded ground truth subbands $\overline{A}_{\nu,6}(n_1,n_2)$, see equation \ref{eq:gt_threshold}, to get the approximations for the wavefront set $D_{\nu,6} (n)$. These act as the training inputs for $\mathcal{N}_2$.

The corresponding ground truth values are computed from the ground truth phantoms, where we have the full wavefront set information available in all of the subbands. We simply take the absolute value, morphologically open, and threshold the data to a binary form. These serve as the ground truth values in the training.

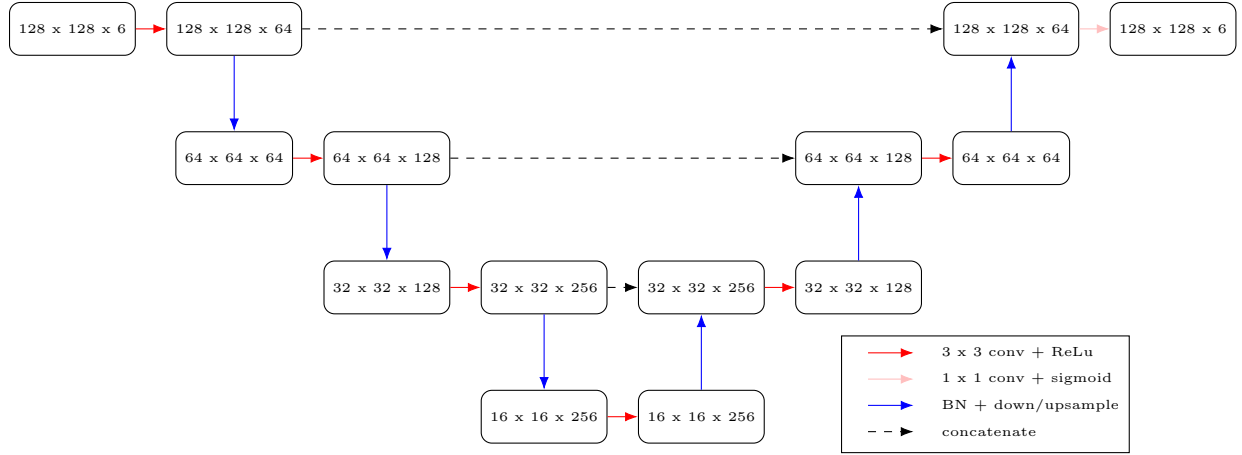
\begin{figure*}[h!]
\input{architecture1}
\caption{Network architecture of $\mathcal{N}_2$.}
\label{fig:nn2_arch}
\end{figure*}

For the neural network $\mathcal{N}_2$, we use a similar architecture as for network $\mathcal{N}_1$, however, here batch normalization is not used in the training (see Figure \ref{fig:nn2_arch}). Otherwise, the training scheme and use of loss function were identical to that of network $\mathcal{N}_1$. See section \ref{sec:learning_the_binary_mask} for a more detailed explanation. Similarly, the training datasets were of size $5000 \times 128 \times 128 \times 6$, from which we used 500 samples for validation.

\subsection{Boundary estimate}
\begin{figure}[h!]
\centering
\begin{subfigure}{.5\textwidth}
  \centering
  \includegraphics[width=45mm]{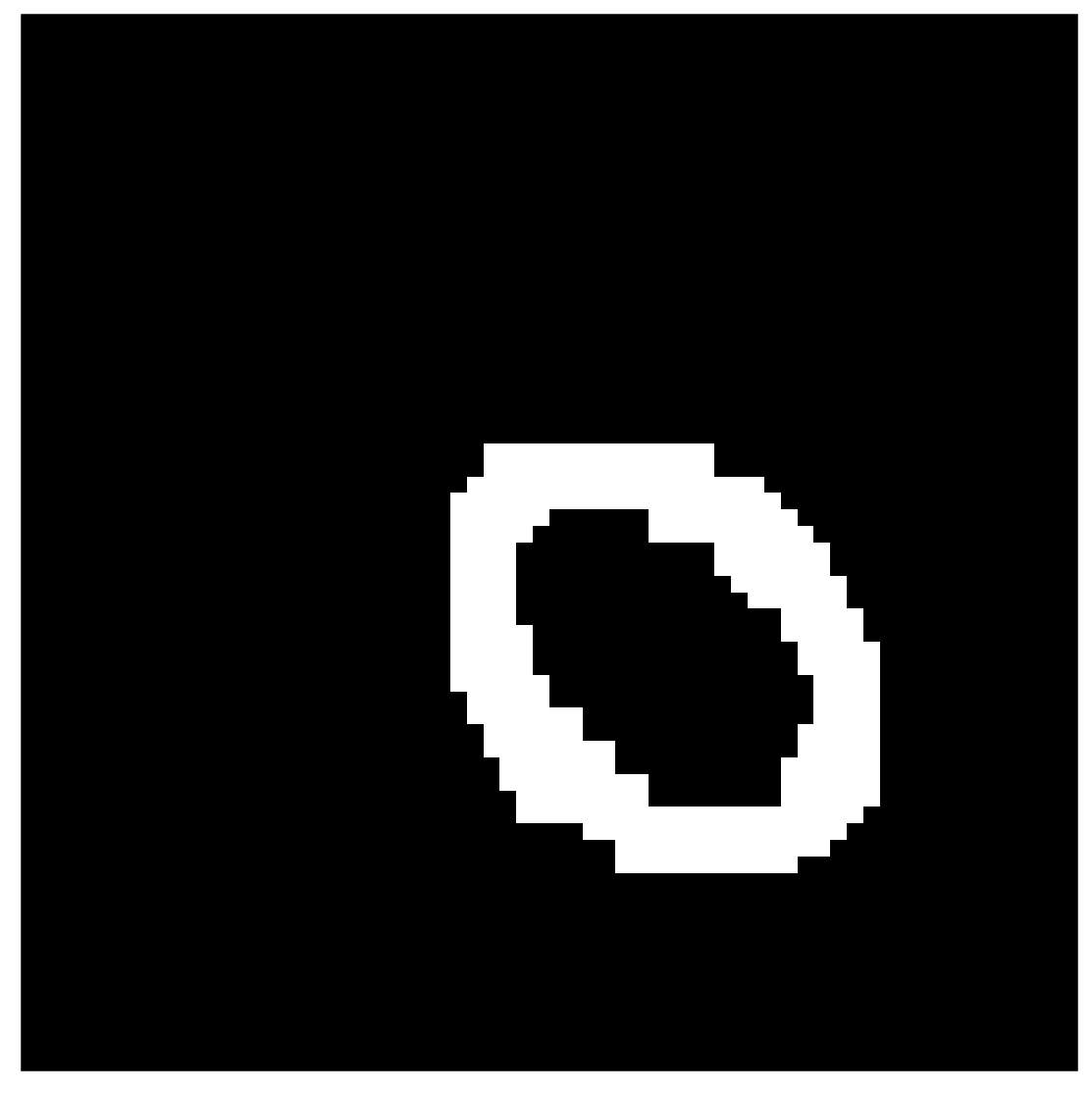}
  \caption{}
  \label{fig:singular_support}
\end{subfigure}%
\begin{subfigure}{.5\textwidth}
  \centering
  \includegraphics[width=45mm]{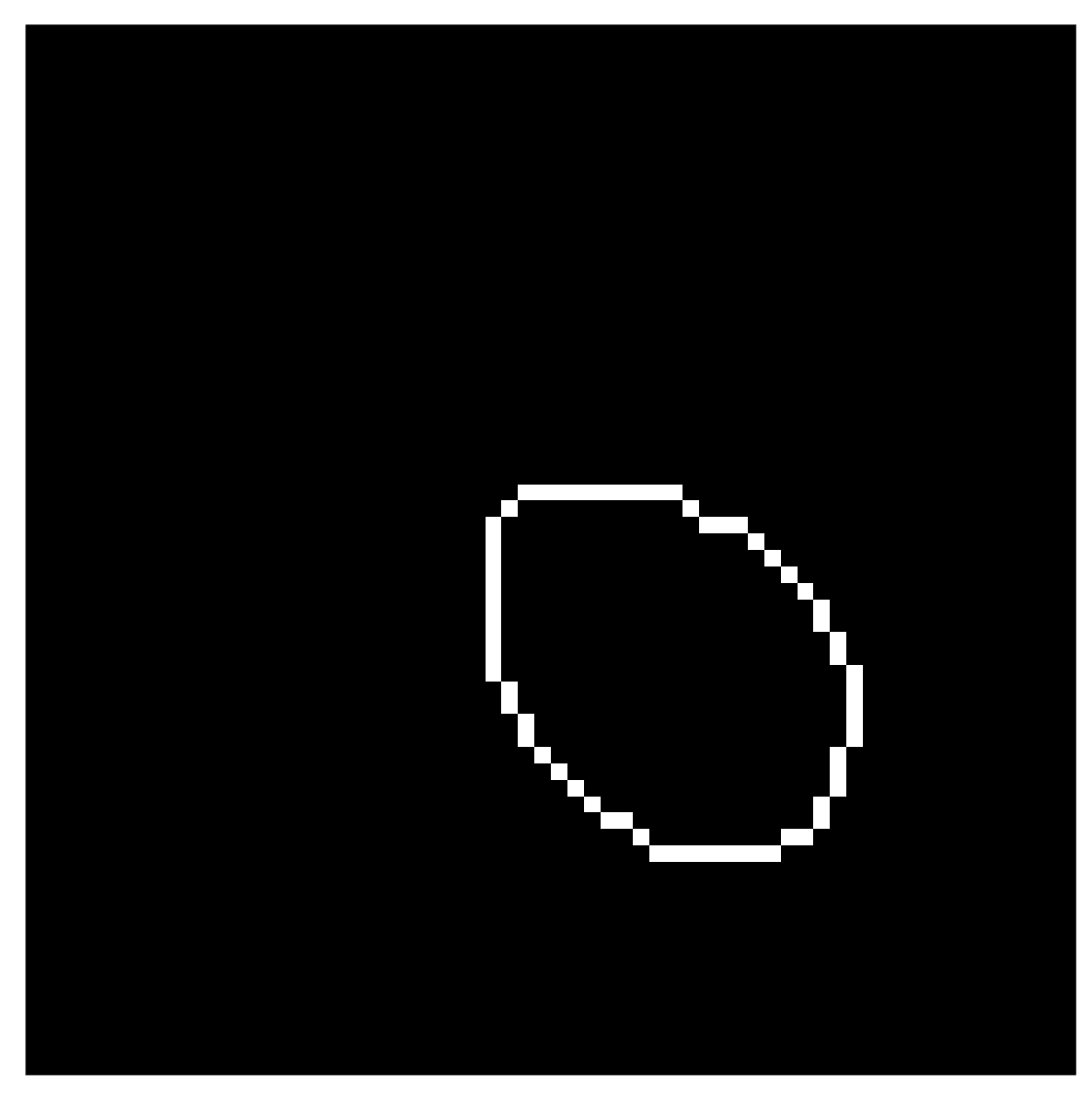}
  \caption{}
  \label{fig:skeleton}
\end{subfigure}
\caption{(a) The singular support $\texttt{singsupp} ( \widehat{D}_{\nu,6})$ computed from the predicted full wavefront set $ \widehat{D}_{\nu,6}$ and (b) its boundary estimate.}
\label{fig:ss_and_skeleton}
\end{figure}

Now, we can combine the extended wavefront set $\widehat{D}_{\nu,6}(n_1,n_2)$, that is the output of network $\mathcal{N}_2$, to form the singular support (see section \ref{section:wst}). That is, we take the maximum of all six binary subbands:

$$\texttt{singsupp} ( \widehat{D}_{\nu,6}) =  \max \Big(\widehat{D}_{(\overline{L}H,6)}, \widehat{D}_{(\overline{H}H,6)},  \widehat{D}_{(\overline{H}L,6)},  \widehat{D}_{(HL,6)},  \widehat{D}_{(HH,6)},  \widehat{D}_{(LH,6)} \Big).$$
See Figure \ref{fig:singular_support} for the full singular support.

Then, we can compute the boundary estimate of the singular support using the morphological skeleton operation described in equation \eqref{morph_skeleton} of section \ref{section:morphological_operations}. See Figure \ref{fig:skeleton}.
Finally, this learned boundary estimate can be used as an overlay for the PDFP reconstruction. See section \ref{section:results} for final results. 



\clearpage
\section{Results} \label{section:results}

\subsection{Reconstructions in the $xz$-plane} \label{section:results2d}
\begin{figure}[b]
    \centering
    \includegraphics[width=0.4\textwidth]{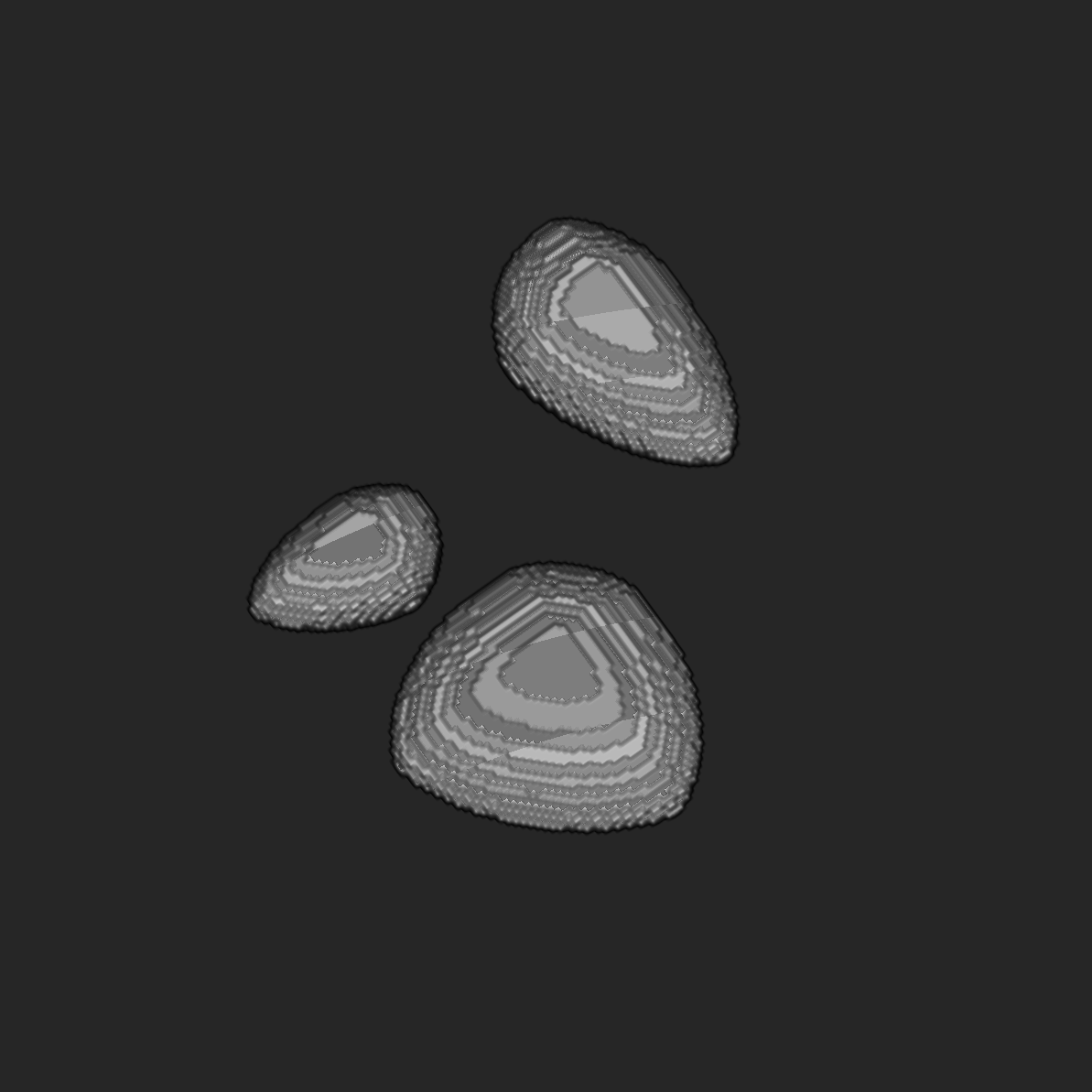}
    \caption{A computational 3D phantom of size $128 \times 128 \times 128$: three $L^p$ -balls, where $p=1.5$, with a constant value of 1, in a 0-valued background.}
    \label{fig:ellipsoid_phantom}
\end{figure}

For testing our proposed method, we created a three-dimensional phantom consisting of three $L^p$ -balls, where $p=1.5$, in a constant background (see Figure \ref{fig:ellipsoid_phantom}). First, we simulated X-ray measurements of the phantom using a parallel-beam imaging geometry of a 40-degree opening angle with 50 projection images. We computed the PDFP reconstructions slice-by slice, separately for each $xz$-plane, treating each slice as an independent 2D tomographic reconstruction problem. The complex wavelet coefficients used for regularization were computed using Kingsbury Q-shift filters. Following the workflow explained in Figure \ref{fig:flowchart}, we learn the boundary estimate for each $xz$-slice.
The learned boundary estimate is then overlayed on the PDFP reconstruction to indicate the extent of boundaries of the stretched features in the reconstruction. 

\begin{figure}[b]
\centering
\begin{subfigure}{.3\textwidth}
  \centering
  \includegraphics[width=43mm]{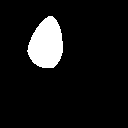}
  \label{fig:og97}
\end{subfigure}%
\begin{subfigure}{.3\textwidth}
  \centering
  \includegraphics[width=43mm]{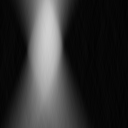}
  \label{fig:ufbp97}
\end{subfigure}%
\begin{subfigure}{.3\textwidth}
  \centering
  \includegraphics[width=43mm]{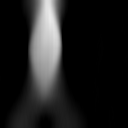}
  \label{fig:pdfp97}
\end{subfigure}%

\vspace{0.5cm}
\begin{subfigure}{.3\textwidth}
  \centering
  \includegraphics[width=43mm]{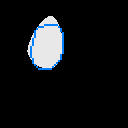}
  \caption{}
  \label{fig:phantomskeleton97}
\end{subfigure}%
\begin{subfigure}{.3\textwidth}
  \centering
  \includegraphics[width=43mm]{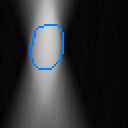}
  \caption{}
  \label{fig:ufbpskeleton97}
\end{subfigure}%
\begin{subfigure}{.3\textwidth}
  \centering
  \includegraphics[width=43mm]{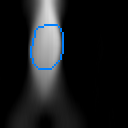}
  \caption{}
  \label{fig:pdfpskeleton97}
\end{subfigure}%
\caption{Comparisons of the $xz$-slice, where $y=57$, of (a) the test phantom, (b) the tomosynthesis reconstruction, and (c) the PDFP reconstruction with complex wavelet regularization. On the bottom row, the learned boundary estimate from the PDFP reconstruction is added to each image as an overlay for comparison purposes. Imaging geometry: parallel beam, 40-degree opening angle (70-110), 50 projections.}
\label{fig:2dreco}
\end{figure}

\begin{figure}[h]
\centering
\begin{subfigure}{.08\textwidth}
   Slice \\$y=22$
\end{subfigure}
\begin{subfigure}{.22\textwidth}
  \centering
  \includegraphics[width=\textwidth]{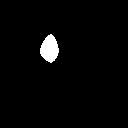}
\end{subfigure}
\begin{subfigure}{.22\textwidth}
  \centering
  \includegraphics[width=\textwidth]{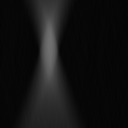}
\end{subfigure}
\begin{subfigure}{.22\textwidth}
  \centering
  \includegraphics[width=\textwidth]{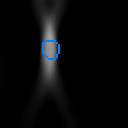}
\end{subfigure}
\begin{subfigure}{.22\textwidth}
  \centering
  \includegraphics[width=\textwidth]{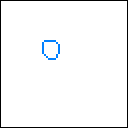}
\end{subfigure}%

\vspace{0.2cm}
\centering
\begin{subfigure}{.08\textwidth}
   $y=61$
\end{subfigure}
\begin{subfigure}{.22\textwidth}
  \centering
  \includegraphics[width=\textwidth]{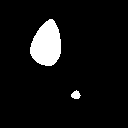}
\end{subfigure}
\begin{subfigure}{.22\textwidth}
  \centering
  \includegraphics[width=\textwidth]{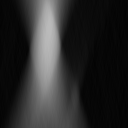}
\end{subfigure}
\begin{subfigure}{.22\textwidth}
  \centering
  \includegraphics[width=\textwidth]{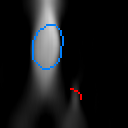}
\end{subfigure}
\begin{subfigure}{.22\textwidth}
  \centering
  \includegraphics[width=\textwidth]{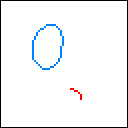}
\end{subfigure}%

\vspace{0.2cm}
\centering
\begin{subfigure}{.08\textwidth}
   $y=84$
\end{subfigure}
\begin{subfigure}{.22\textwidth}
  \centering
  \includegraphics[width=\textwidth]{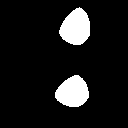}
\end{subfigure}
\begin{subfigure}{.22\textwidth}
  \centering
  \includegraphics[width=\textwidth]{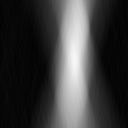}
\end{subfigure}
\begin{subfigure}{.22\textwidth}
  \centering
  \includegraphics[width=\textwidth]{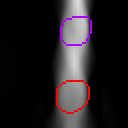}
\end{subfigure}
\begin{subfigure}{.22\textwidth}
  \centering
  \includegraphics[width=\textwidth]{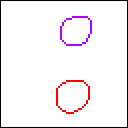}
\end{subfigure}%

\vspace{0.2cm}
\centering
\begin{subfigure}{.08\textwidth}
   $y=94$
\end{subfigure}
\begin{subfigure}{.22\textwidth}
  \centering
  \includegraphics[width=\textwidth]{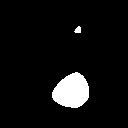}
\end{subfigure}
\begin{subfigure}{.22\textwidth}
  \centering
  \includegraphics[width=\textwidth]{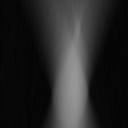}
\end{subfigure}
\begin{subfigure}{.22\textwidth}
  \centering
  \includegraphics[width=\textwidth]{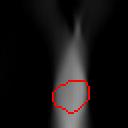}
\end{subfigure}
\begin{subfigure}{.22\textwidth}
  \centering
  \includegraphics[width=\textwidth]{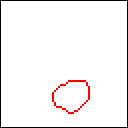}
\end{subfigure}%

\vspace{0.2cm}
\centering
\begin{subfigure}{.08\textwidth}
   $y=113$
\end{subfigure}
\begin{subfigure}{.22\textwidth}
  \centering
  \includegraphics[width=\textwidth]{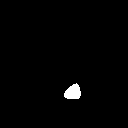}
  \caption{}
\end{subfigure}
\begin{subfigure}{.22\textwidth}
  \centering
  \includegraphics[width=\textwidth]{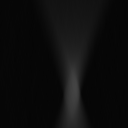}
  \caption{}
\end{subfigure}
\begin{subfigure}{.22\textwidth}
  \centering
  \includegraphics[width=\textwidth]{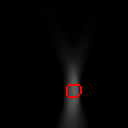}
  \caption{}
\end{subfigure}
\begin{subfigure}{.22\textwidth}
  \centering
  \includegraphics[width=\textwidth]{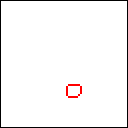}
  \caption{}
\end{subfigure}%
\caption{Different $xz$-slices through the three-dimensional test phantom shown in Figure \ref{fig:ellipsoid_phantom}. In the model, $xz$-slices are vertical (perpendicular to the detector surface). In our simplified assumption of parallel-beam geometry, there is an independent limited-angle tomography problem restricted to each $xz$-slice. (a) Ground truth slice. (b) Tomosynthesis reconstruction. (c) PDFP reconstruction with learned boundary estimate. (d) Boundary estimate curve. Compare the boundary curves to the true boundaries of features shown in column (a).}
\label{fig:3dXZreco}
\end{figure}

In Figure \ref{fig:2dreco}, a reconstructed $xz$-slice with the learned boundary estimate is compared to the ground truth $xz$-slice of the test phantom. Also, the tomosynthesis reconstruction is shown for comparison. 
The results of additional $xz$-slices are shown in Figure \ref{fig:3dXZreco}. To measure how well the learned boundary estimate matches the true boundary, we have computed the dice similarity coefficient (DSC) of segmentations computed from the learned boundary estimate ($X_z$) and compared those to the ground truth slice ($Y_z$):
$$\textrm{DSC} = \frac{2 |X_z \cap Y_z|}{|X_z| + |Y_z|},$$
where $|X_z|, |Y_z|$ are the cardinalities, that is, the number of elements in the sets. The resulting DSC values are presented in table \ref{table:dice_sim_coeffs}.

\subsection{Reconstructions in the $xy$-plane} \label{section:results3d}

\begin{figure}[h]
\centering
\begin{subfigure}{.08\textwidth}
   Slice \\$z=7$
\end{subfigure}
\begin{subfigure}{.22\textwidth}
  \centering
  \includegraphics[width=\textwidth]{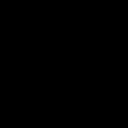}
\end{subfigure}
\begin{subfigure}{.22\textwidth}
  \centering
  \includegraphics[width=\textwidth]{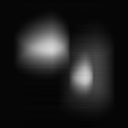}
\end{subfigure}
\begin{subfigure}{.22\textwidth}
  \centering
  \includegraphics[width=\textwidth]{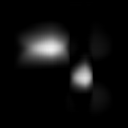}
\end{subfigure}
\begin{subfigure}{.22\textwidth}
  \centering
  \includegraphics[width=\textwidth]{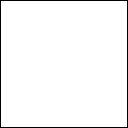}
\end{subfigure}%

\vspace{0.2cm}
\centering
\begin{subfigure}{.08\textwidth}
  $z=35$
\end{subfigure}
\begin{subfigure}{.22\textwidth}
  \centering
  \includegraphics[width=\textwidth]{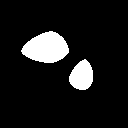}
\end{subfigure}
\begin{subfigure}{.22\textwidth}
  \centering
  \includegraphics[width=\textwidth]{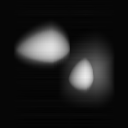}
\end{subfigure}
\begin{subfigure}{.22\textwidth}
  \centering
  \includegraphics[width=\textwidth]{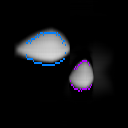}
\end{subfigure}
\begin{subfigure}{.22\textwidth}
  \centering
  \includegraphics[width=\textwidth]{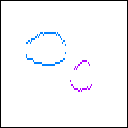}
\end{subfigure}

\vspace{0.2cm}
\centering
\begin{subfigure}{.08\textwidth}
  $z=57$
\end{subfigure}
\begin{subfigure}{.22\textwidth}
  \centering
  \includegraphics[width=\textwidth]{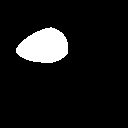}
\end{subfigure}
\begin{subfigure}{.22\textwidth}
  \centering
  \includegraphics[width=\textwidth]{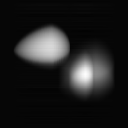}
\end{subfigure}
\begin{subfigure}{.22\textwidth}
  \centering
  \includegraphics[width=\textwidth]{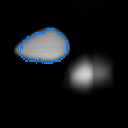}
\end{subfigure}
\begin{subfigure}{.22\textwidth}
  \centering
  \includegraphics[width=\textwidth]{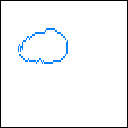}
\end{subfigure}

\vspace{0.2cm}
\centering
\begin{subfigure}{.08\textwidth}
  $z=94$
\end{subfigure}
\begin{subfigure}{.22\textwidth}
  \centering
  \includegraphics[width=\textwidth]{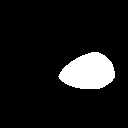}
\end{subfigure}
\begin{subfigure}{.22\textwidth}
  \centering
  \includegraphics[width=\textwidth]{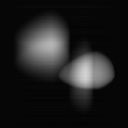}
\end{subfigure}
\begin{subfigure}{.22\textwidth}
  \centering
  \includegraphics[width=\textwidth]{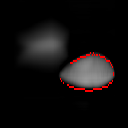}
\end{subfigure}
\begin{subfigure}{.22\textwidth}
  \centering
  \includegraphics[width=\textwidth]{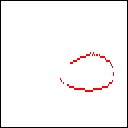}
\end{subfigure}

\vspace{0.2cm}
\begin{subfigure}{.08\textwidth}
  $z=117$
\end{subfigure}
\begin{subfigure}{.22\textwidth}
  \centering
  \includegraphics[width=\textwidth]{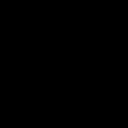}
  \caption{}
\end{subfigure}
\begin{subfigure}{.22\textwidth}
  \centering
  \includegraphics[width=\textwidth]{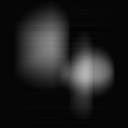}
  \caption{}
\end{subfigure}
\begin{subfigure}{.22\textwidth}
  \centering
  \includegraphics[width=\textwidth]{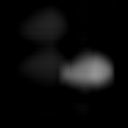}
  \caption{}
\end{subfigure}
\begin{subfigure}{.22\textwidth}
  \centering
  \includegraphics[width=\textwidth]{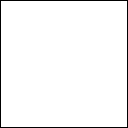}
  \caption{}
\end{subfigure}
\caption{Different $xy$-slices through the three-dimensional test phantom of Figure \ref{fig:ellipsoid_phantom} and through 3D-volumes achieved by stacking 2D-reconstructions from $xz$-planes. In the model, $xy$-slices are parallel to the detector surface. (a) Ground truth slice. (b) Tomosynthesis. (c) PDFP reconstruction with learned boundary estimate. (d) Boundary estimate curve. Note that one can deduce from the boundary curves that the gray features in the top and bottom images of column (c) are artefacts. 
}
\label{fig:3dXYreco}
\end{figure}

For a three-dimensional phantom, after independently computing PDFP reconstructions and the boundary estimates for each of the $xz$-slices separately (see section \ref{section:results2d}), we can stack the reconstruction results back together to form a volume. Then, we can slice the reconstructed volume in the $xy$-direction. 
We present the results as $xy$-slices with a selection of different $z$-values. Reconstruction results of several $xy$-slices of the phantom are shown in Figure \ref{fig:3dXYreco}. 

\begin{table}[h!]
\begin{center}
\begin{tabular}{|c c|} 
 \hline
 xz-slice (y)& DSC \\ [0.5ex] 
 \hline\hline
 22 & 0.68966 \\ 
 \hline
 57 & 0.86774 \\
 \hline
 61 & 0.85950 \\
 \hline
 84 & 0.83652 \\
 \hline
 94 & 0.79417 \\
 \hline
 113 & 0.71338 \\ [1ex] 
 \hline
\end{tabular}
\end{center}
\caption{Dice similarity coefficients for segmentations computed from the learned boundary estimate and compared to the ground truth phantom in the $xz$-plane. The chosen slices correspond to those shown in Figures \ref{fig:2dreco}, \ref{fig:3dXZreco}.}
\label{table:dice_sim_coeffs}
\end{table}




\clearpage

\clearpage

\clearpage

\section{Conclusion} \label{section:discussion}


We introduced a new method for estimating the wavefront set of an X-ray coefficient function, given limited-angle tomography data. We found that detecting the wavefront set by thresholding complex wavelet coefficients was difficult to make automatic. However, we were able to train a neural network to find a good approximation. In addition, our idea of using geometric {\it a priori} information in learning to continue the visible part of the wavefront set to the invisible part was successful. 


We trained our method with simulated elliptical targets in $2$D. Our data was severely limited with a 40-degree angle-of-view. We found that the method generalizes successfully to 2D problems arising from slices of simulated $L^p$-ball targets in $\R^3$. Namely, the extent of objects in the direction of the $z$-axis was rather accurately recovered. However, the method did have difficulties near places where the boundaries of $L^p$-balls had high curvature. We also noted that if some inclusions were much smaller than others (or of less contrast), the method might not detect them.

Our computational example suggests that our method can recover boundary curves with unprecedented accuracy for smooth inclusions. We expect that to be useful, for instance, in detecting bubbles when inspecting welds. However, for DBT, it remains unclear what happens with erratic boundaries of tumors. 

The natural next step in this research direction is to fix an application and apply the method to measured data. The generalization to fan-beam geometry is straight-forward and in principle, there should be no issues in extending our method to 3D with cone-beam imaging geometry.

\section*{Acknowledgements}
Siiri Rautio was supported by a University of Helsinki -funded doctoral researcher position. Rashmi Murthy was funded by the Future Makers project AIDMEI, funded by the Jane and Aatos Erkko Foundation and Technology Industries of Finland Centennial Foundation. Tatiana A. Bubba was supported by the Academy of Finland (310822) and by the Royal Society through the Newton International Fellowship grant n.~NIF\textbackslash R1\textbackslash 201695. Matti Lassas and Samuli Siltanen are supported by the Finnish Centre of Excellence in Inverse Modelling and Imaging, 2018-2025, decision numbers 312339 and 336797, Academy of Finland grants 284715 and  312110.

\clearpage
\bibliographystyle{plain}

\bibliography{DBT_candywrap}

\end{document}

%% file: architecture1.tex
\begin{center}
\begin{tikzpicture} 

    \node[module=0.5cm,font=\fontsize{4}{4}\selectfont] (I1) {128 x 128 x 6};
    \node[module=0.5cm,font=\fontsize{4}{4}\selectfont,right= 4mm of I1] (I2) {128 x 128 x 64};
    
    \node[module=0.5cm,font=\fontsize{4}{4}\selectfont, below=of I2] (I3) {64 x 64 x 64};
    \node[module=0.5cm,font=\fontsize{4}{4}\selectfont, right= 4mm of I3] (I4) {64 x 64 x 128};
    
    \node[module=0.5cm,font=\fontsize{4}{4}\selectfont, below=of I4] (I5) {32 x 32 x 128};
    \node[module=0.5cm,font=\fontsize{4}{4}\selectfont, right= 4mm of I5] (I6) {32 x 32 x 256};
    
    \node[module=0.5cm, font=\fontsize{4}{4}\selectfont,below=of I6] (I7) {16 x 16 x 256};
    \node[module=0.5cm, font=\fontsize{4}{4}\selectfont,right= 4mm of I7] (I8) {16 x 16 x 256};
    
    \node[module=0.5cm, font=\fontsize{4}{4}\selectfont,above=of I8] (I9) {32 x 32 x 256};
    \node[module=0.5cm, font=\fontsize{4}{4}\selectfont,right= 4mm of I9] (I10) {32 x 32 x 128};
    
    \node[module=0.5cm, font=\fontsize{4}{4}\selectfont,above=of I10] (I11) {64 x 64 x 128};
    \node[module=0.5cm, font=\fontsize{4}{4}\selectfont,right= 4mm of I11] (I12) {64 x 64 x 64};
    
    \node[module=0.5cm, font=\fontsize{4}{4}\selectfont,above=of I12] (I13) {128 x 128 x 64};
    \node[module=0.5cm, font=\fontsize{4}{4}\selectfont,right= 4mm of I13] (I14) {128 x 128 x 6};
    
    \node[below= 21mm of I11] (L1) {};
    \node[right= 6mm of L1] (L2) {};
    \node[right= 0mm of L2, font=\fontsize{4}{4}\selectfont] (L3) {3 x 3 conv + ReLu};
    
    \node[below= 1mm of L1] (L4) {};
    \node[right= 6mm of L4] (L5) {};
    \node[right= 0mm of L5, font=\fontsize{4}{4}\selectfont] (L6) {1 x 1 conv + sigmoid};
    
    \node[below= 1mm of L4] (L7) {};
    \node[right= 6mm of L7] (L8) {};
    \node[right= 0mm of L8, font=\fontsize{4}{4}\selectfont] (L9) {BN + down/upsample};
    
    \node[below= 1mm of L7] (L10) {};
    \node[right= 6mm of L10] (L11) {};
    \node[right= 0mm of L11, font=\fontsize{4}{4}\selectfont] (L12) {concatenate};
    
    \node[fit=(L1)(L6) (L12), draw, inner sep=1mm] (fit1) {};
    
    \draw[->,red] (I1)--(I2);
    \draw[->,blue] (I2)--(I3);
    \draw[->,red] (I3)--(I4);
    \draw[->,blue] (I4)--(I5);
    \draw[->,red] (I5)--(I6);
    \draw[->,blue] (I6)--(I7);
    \draw[->,red] (I7)--(I8);
    \draw[->,blue] (I8)--(I9);
    \draw[->,red] (I9)--(I10);
    \draw[->,blue] (I10)--(I11);
    \draw[->,red] (I11)--(I12);
    \draw[->,blue] (I12)--(I13);
    \draw[->,pink] (I13)--(I14);
    
    \draw[->,dashed] (I4)--(I11);
    \draw[->,dashed] (I2)--(I13);
    \draw[->,dashed] (I6)--(I9);
    
    \draw[->,red] (L1)--(L2);
    \draw[->,pink] (L4)--(L5);
    \draw[->,blue] (L7)--(L8);
    \draw[->,dashed] (L10)--(L11);
    
\end{tikzpicture}
\end{center}